\documentstyle[psfig]{mn}
\newcommand{\HI}{{\rm H\,\scriptstyle I}}
\newcommand{\HII}{{\rm H\,\scriptstyle II}}

\title{Scaling and correlation analysis of galactic images}

\author[P.~Frick et al.]
       {P.~Frick,$^{1,2}$ R.~Beck,$^2$ E.M.~Berkhuijsen$^2$ and
        I.~Patrickeyev$^{1,2}\thanks{E-mail: frick@icmm.ru (PF);
                rbeck@mpifr-bonn.mpg.de (RB);\newline
                eberkhuijsen@mpifr-bonn.mpg.de (EMB); pat@icmm.ru (IP)} $\\
$^1$ Institute of Continuous Media Mechanics, Korolyov
        str.~1, 614061 Perm, Russia \\
$^2$ Max-Planck-Institut f\"ur Radioastronomie, Auf dem H\"ugel 69,
        D-53121 Bonn, Germany}

\date{Accepted 2001 July 5.
      Received 2001 June 13;
      in original form 2001 January 31}
\pagerange{\pageref{firstpage}--\pageref{lastpage}}

\begin{document}

\label{firstpage}
\maketitle

\begin{abstract}
Different scaling and autocorrelation characteristics and their
application to astronomical images are discussed: the structure
function, the autocorrelation function, Fourier spectra and wavelet
spectra. The choice of the mathematical tool is of great importance for
the scaling analysis of images. The structure function, for example,
cannot resolve scales which are close to the dominating
large-scale structures and can lead to the wrong interpretation
that a continuous range of scales with a power law exists. The
traditional Fourier technique, applied to real data, gives very spiky
spectra, in which the separation of real maxima and high harmonics can
be difficult. We recommend as the optimal tool the {\it wavelet spectrum}
with a suitable choice of the analysing wavelet. We
introduce the {\it wavelet cross-correlation function} which
enables to study the correlation between images as a function of
scale. The cross-correlation coefficient strongly depends on the scale.
The classical cross-correlation coefficient can be misleading if a
bright, extended central region or an extended disk exists in the
galactic images.

An analysis of the scaling and cross-correlation characteristics of 9
optical and radio maps of the nearby spiral galaxy NGC\,6946 is
presented. The wavelet spectra allow to separate structures on
different scales like spiral arms and diffuse extended emission.
Only the images of thermal radio emission and H$\alpha$ emission give
indications of 3-dimensional Kolmogorov-type turbulence on the
smallest resolved scales (160--800~pc).
The cross-correlations between the images of NGC\,6946 show strong
similarities between the images of total radio emission, red light and
mid-infrared dust emission on all scales. The best correlation
is found between total radio emission and
dust emission. Thermal radio continuum and H$\alpha$ emission
are best correlated on a scale of about $1\arcmin \simeq 1.6$~kpc,
the typical width of a spiral arm. On a similar scale, the images of
polarised radio and H$\alpha$ emission are {\it anticorrelated},
which remains undetected with classical cross-correlation analysis.
\end{abstract}



\section{Introduction}

The increasing resolution of optical, radio, infrared and X-ray
telescopes over the past decades has greatly improved the details
visible in astronomical objects in many wavelength ranges. The
large variety in the structure of external galaxies, which have
bars, bright central regions, spiral arms with clumpy
star-forming regions, long known from optical photographs, can
now also be identified in radio, X-ray and IR continuum, and in
spectral lines of ionised ($\HII$), atomic ($\HI$) and molecular
(CO) gas components. How these different constituents in a galaxy
are related is a question to which continuous study has been devoted.

The detection of similar structures in widely separated energy ranges
gives important information on physical processes and their interplay.
For example, the degree of correlation between the H$\alpha$ emission
and the free-free emission in radio continuum tells us about the amount
of absorption in the optical range. The similarity of far-infrared and
radio continuum emission components (e.g. in the nearby spiral galaxy
M\,31) indicates that magnetic fields are not anchored in the warm
medium, but in cool gas clouds (Hoernes, Berkhuijsen \& Xu 1998). Also
emission in the CO line is correlated with radio continuum emission in
spiral arms of M\,31, but it is anticorrelated with the polarised radio
emission (Berkhuijsen, Bajaja \& Beck 1993).

Anticorrelations are possibly even more interesting than correlations.
In the spiral galaxy NGC\,6946 a striking anticorrelation between the
optical spiral arms and the ``magnetic arms'' seen in radio polarisation
was discovered by Beck \& Hoernes (1996). Although obvious to the eye,
the analysis of this phenomenon needs sophisticated techniques.
Based on wavelet analysis, Frick et al. (2000) showed that the ``magnetic
arms'' are phase-shifted images of the optical arms. Dynamo action is
able to generate such structures (Rohde, Beck \& Elstner 1999).

The existence of a correlation, anticorrelation or non-correlation has
important consequences for the interpretation of observable quantities
which emerge from a combination of physical quantities. For example,
Faraday rotation ($RM$) of polarised radio waves is due to the product
of electron density ($n_{\rm e}$) and the magnetic field component along
the line of sight ($B_{\parallel}$), averaged along the line of sight.
Knowledge of $<n_{\rm e}>$ (e.g. from dispersion measures of pulsars)
is not sufficient to determine the strength of $<B_{\parallel}>$ unless
it is known whether $n_{\rm e}$ and $B_{\parallel}$ are correlated,
anticorrelated or non-correlated. If the anticorrelation observed on
the scale of spiral arms holds also on small scales, the field
strengths derived from Faraday rotation data of pulsars are too small
(Beck 2001).

Relations between images are generally analysed with the pixel-to-pixel
correlation function. However, this method gives little information in
the case of an anticorrelation on the scale of spiral arms, like in
NGC\,6946, or when the diffuse emission on larger scales (e.g. in total
radio emission) has no counterpart in the other image (e.g. polarised
radio emission).

Another interesting subject is turbulence in the interstellar medium of
galaxies, which has been studied by analysing structure functions of
appropriate images. Observations of radio-wave scintillations in the
Milky Way have revealed the existence of density irregularities in the
diffuse ionised gas with scale sizes of $\ge 10^{18}$~cm (0.3~pc)
down to $10^7$~cm. The spectrum of the irregularities is consistent with the
spectrum of three-dimensional Kolmogorov turbulence (Spangler 1999).
A spectrum with the same slope was found by Minter \& Spangler (1996)
for both the turbulent fluctuations in electron density and in
Faraday rotation measure observed in the same area in the Milky Way.
The outer scale of these fluctuations is 4~pc. For larger scales up to
about 80~pc Minter \& Spangler found a flatter spectrum. As turbulent
scales larger than about 100~pc are difficult to observe in our Galaxy,
it is still unknown whether the Kolmogorov spectrum extends to large
scales. The analysis of images of external galaxies may give us this
information.

In laboratory experiments extremely long data series are used to get
reliable statistics for the slopes of structure functions of high order
in the inertial range (Monin \& Yaglom 1971, 1975; Frisch 1995). The
analysis of turbulence in astronomical objects requires observational
data of very high quality and high angular resolution. The excellent
data set on the external galaxy NGC\,6946 seems well suited for a study
of turbulence on large scales. In a first attempt Beck, Berkhuijsen \&
Uyan\i ker (1999) analysed the high-resolution radio continuum maps of
this galaxy using the structure function. They found spectra consistent
with two-dimensional turbulence in the radio continuum emission on
scales between $0\farcm 5$ (twice the beam size) and $5\farcm 0$. In
this paper, however, we show that in the case of images with pronounced
structures on scales less than 20$\times$ the size of the telescope
beam, like the bright central area and spiral arms covering most of the
available images, the structure function is not the best analysing tool
because it does not well resolve these scales. The lack of spectral
resolution and the relatively poor statistics (i.e. a small ratio of
image size to beam size)  influence the slope of the derived structure
function, making the interpretation difficult. As astronomical
observations are limited in resolution and extent, the mathematical
tools used for their analysis have to be chosen carefully and the
results need a critical evaluation.

During the last decade a new mathematical tool has been developed,
{\it wavelet analysis\/} which enables the detection of structures of
different scales in data sets. First applications to astronomical
objects showed that wavelets are effective in time series analysis
(Foster 1996; Frick et al. 1997a,b), denoising (Tenorio et al. 1999;
Chen et al. 2000) and structure detection (Frick et al. 2000).

We have applied two-dimensional wavelet analysis to images of
NGC\,6946 at various wavelengths. Our goal is not only to detect the
dominant scales in this galaxy, but also to see if wavelets can be
useful to determine the {\it statistical\/} characteristics for {\it
given scales\/} in the maps. The observation of magnetic arms situated
in between the gaseous arms, leading to an absence of classical
cross-correlation between the corresponding images in spite of their
similarity, stimulated this research.

The paper is organized as follows. In Sect.~2 we present the wavelets
as a tool of scaling analysis. Their relationship with spectra and
structure functions is explained in Sect.~3, and how wavelets can be
used for correlations at a given scale is shown in Sect.~4. In the
second part of the paper the wavelet analysis is applied to images of
the nearby galaxy NGC\,6946 (Sect.~5.1). First its spectral
characteristics are discussed in Sect.~5.2, then the correlations
between pairs of images are presented in Sect.~5.3. The results are
discussed in Sect.~6.

\section{Wavelets as a tool for scaling analysis}

Wavelet analysis is based on a space-scale decomposition using the
convolution of the data with a family of self-similar basic functions
that depend on two parameters, scale and location. It can be considered
as a generalization of the Fourier transformation, which uses harmonic
functions as a one-parametric functional basis, characterized by
frequency, or in the case of a space function, by the wavevector $\vec
k$. The wavelet transformation also uses oscillatory functions, but in
contrast to the Fourier transform these functions rapidly decay towards
infinity. The family of functions is generated by dilations and
translations of the mother function, called the analysing wavelet. This
procedure provides self-similarity, which distinguishes the wavelet
technique from the windowed Fourier transformation, where the
frequency, the width of the window and its position are independent
parameters.

We consider the continuous wavelet transform, which in the
two-dimensional case can be written in the form
\begin{equation}
W(a,\vec x) = {1 \over {a^{\kappa}}}\int_{-\infty}^{+\infty}
\int_{-\infty}^{+\infty} f(\vec x')
\psi^* \left( {{\vec x'-\vec x}\over a}\right) d \vec x'.
\label{w_trans}
\end{equation}
Here $\vec x = (x,y)$, $f(\vec x)$ is a two-dimensional function,
for which the Fourier transform exists (i.e. square integrated),
$\psi(\vec x)$ is the analysing wavelet (real or complex, $^*$
indicates the complex conjugation), $a$ is the scale parameter,
and $\kappa$ is a normalization parameter which will be discussed
below.

For later considerations the relation between the wavelet and the Fourier
decomposition will be useful. The 2-D Fourier transform
$\hat f(\vec k)$ of the function $f(\vec x)$ is defined as
\begin{equation}
\hat f(\vec k)  =  \int_{-\infty}^{+\infty}
\int_{-\infty}^{+\infty} f(\vec x) e^{-i\vec k\vec x}d\vec x,
\label{four}
\end{equation}
where $\vec k=(k_x,k_y)$ is the wavevector. Then the inverse Fourier
transform is
\begin{equation}
f(\vec x)  =  {1\over{4\pi^2}}\int_{-\infty}^{+\infty}
\int_{-\infty}^{+\infty} \hat f(\vec k)
e^{i \vec k\vec x} d\vec k
\label{four_in}
\end{equation}
and the wavelet coefficients
(\ref{w_trans}) can be expressed as
\begin{equation}
W(a,\vec x) = {a^{2-\kappa}\over{4\pi^2}}\int_{-\infty}^{+\infty}
\int_{-\infty}^{+\infty} \hat f(\vec k)
\hat \psi^* ( a \vec k) e^{i\vec k\vec x} d \vec k.
\label{w_four}
\end{equation}

We restrict our analysis to the use of {\it isotropic\/} wavelets.
It means that the analysing wavelet is an axisymmetric function
$\psi = \psi (\rho),\  \rho = \sqrt{x^2+y^2}$. The choice of the
wavelet function depends on the data and on the goals of the
analysis. For spectral analysis wavelets with good spectral
resolution (i.e. well localized in Fourier space, or having many
oscillations) are preferable, for local structure recognition a
function, well localized in the physical space, is preferable.
(Note that the spectral resolution $\Delta k$ and the space
resolution $\Delta x$ are strongly related and are restricted by
the uncertainty relation $\Delta x \Delta k \geq 2\pi$.) An
obligatory property of the wavelet is the zero mean value $\int
\int \psi(x,y)dx dy =0$.

In the case under consideration the choice of the wavelet was
determined by the wish to have more independent points for further
structure analysis which led to a simple real isotropic wavelet with
a minimal number of oscillations, known as the {\it Mexican Hat}
\begin{equation}
\psi(\rho) = (2-\rho^2)e^{-\rho^2/2}.
\label{hat}
\end{equation}
We shall refer to this function as MH.

\begin{figure}
\centerline{\psfig{file=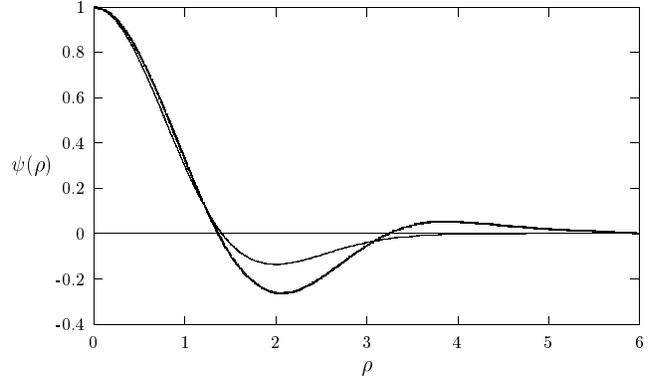,width=8.4truecm,%
       bbllx=80pt,bblly=315pt,bburx=431pt,bbury=527pt}}
\caption{Two isotropic wavelet functions used in this paper:
the MH wavelet (thin line) and the PH wavelet (thick line).}
\label{functions}
\end{figure}

For a better separation of scales (to analyse spectra and to find
the scale of dominant structures) another isotropic wavelet was used
which is defined in Fourier space by the formula
\begin{equation}
\hat \psi(\vec k) = \left\{ \begin{array}
{l@{\qquad : \qquad }l}
\cos^2({\pi\over 2}\log_2{k \over{2\pi}}) & \pi<|\vec k|<4\pi \\
0 & |\vec k|<\pi, |\vec k|>4\pi \ .
\end{array} \right.
\label{myf}
\end{equation}
The function is
localized in Fourier space in a ring with a median radius $2\pi$
and vanishes for $|\vec k|<\pi$ and $|\vec k|>4\pi$. This wavelet
definition provides a relatively good spectral resolution, it
will be referred to as PH (it is our {\it Pet Hat}).
\footnote{This function was introduced in Aurell et al. (1994) for
modeling the two-dimensional turbulence.}
In physical space the PH wavelet is obtained by numerical integration
of (\ref{myf}). Both wavelets MH and PH are shown in Fig.~\ref{functions}.

The wavelet transform (\ref{w_trans}) is unique and reversible
which means that the analysed function $f(x,y)$ can be reconstructed
from its wavelet decomposition. In our analysis we do not need the
inverse transform and do not give the reconstruction formula. (An
extended description of continuum wavelet transform can be found in
e.g. Holschneider (1995) and Torresani (1995).)

\begin{figure*}
\centerline{\psfig{file=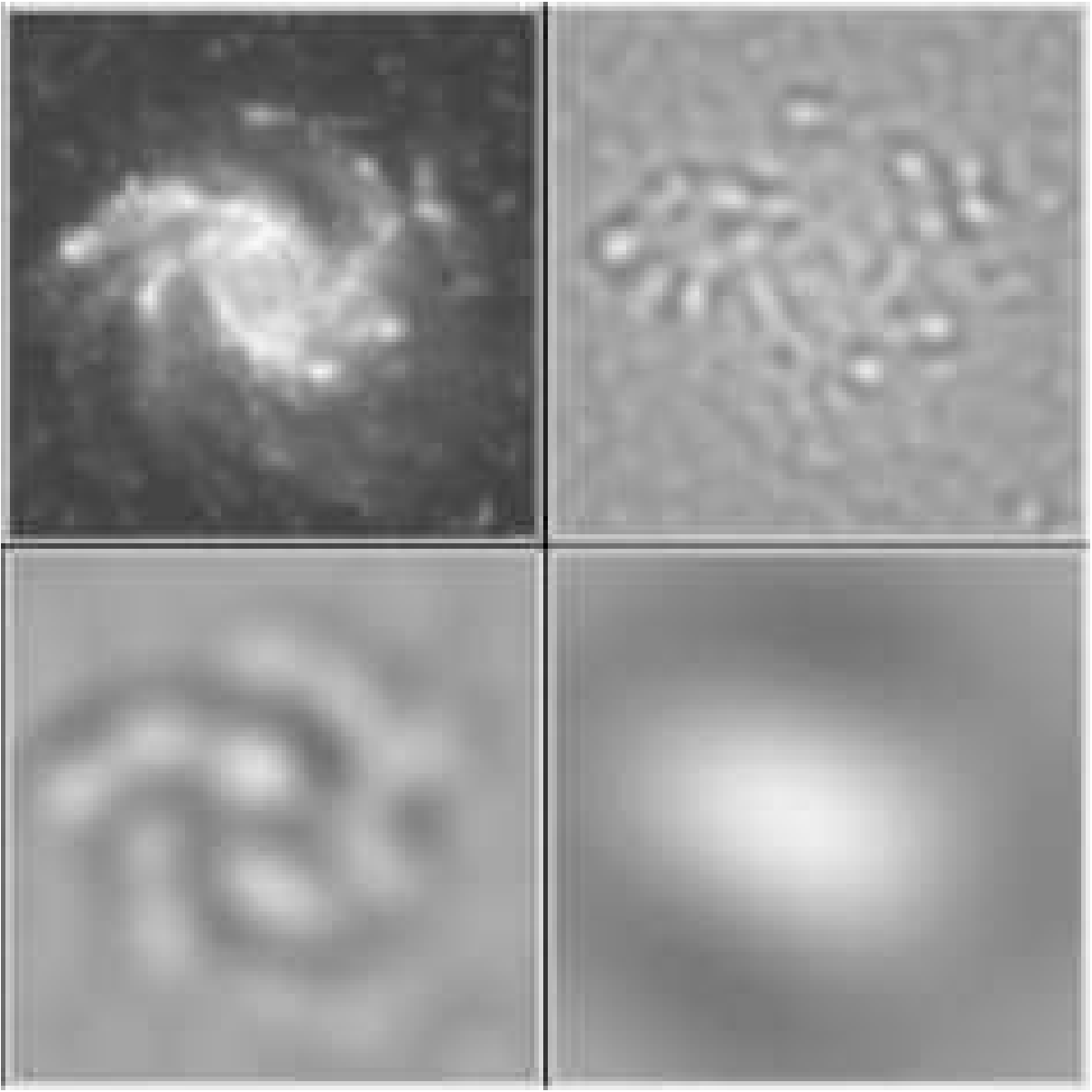,angle=270,width=16truecm,%
       bbllx=14pt,bblly=14pt,bburx=582pt,bbury=582pt}}
\caption{NGC\,6946. Map of the optical emission in red light
(central region subtracted) and its wavelet decompositions for 3
different scales: 0\farcm 5, 1\farcm 4 and 4\arcmin. The size of each
image is $11\arcmin \times 11\arcmin$.}
\label{red_slices}
\end{figure*}

To illustrate how wavelets decompose the image in different scales we
show in Fig.~\ref{red_slices} the optical broadband emission image
of the galaxy NGC\,6946 and its wavelet coefficients $W(a,\vec{x})$ for
three different scales $a$. For this example the PH wavelet was used.

\section{Spectra and structure functions}

In the studies of scaling properties of turbulence a commonly used
characteristic of a turbulent field is the {\it spectral energy
density} $E(k)$ which includes the energy $F(\vec k)=| \hat f(\vec
k)|^2$ of all Fourier harmonics with wavenumbers $\vec k$, for which
$|\vec k|=k$
\begin{equation}
E(k) = \int_{|\vec k|=k} F(\vec k)d \vec k \ .
\label{en_den}
\end{equation}
The spectral energy is related to the {\it autocorrelation\/} function
\begin{equation}
C(\vec l) = \int_{-\infty}^{+\infty} \int_{-\infty}^{+\infty}
f(\vec x)f(\vec l - \vec x) d\vec x
\label{autocor}
\end{equation}
by its Fourier transform
\begin{equation}
F(\vec k) = \hat C(\vec k) = \int_{-\infty}^{+\infty}
\int_{-\infty}^{+\infty} C(\vec l) e^{-i\vec k\vec l} d\vec l \ .
\label{hinchin}
\end{equation}
The vector $\vec l$ defines the shift of the image in the convolution
(\ref{autocor}) and is the scale parameter. In the important case of
isotropic turbulence the autocorrelation function depends only on the
distance between two points  $C(\vec l)=C(l)$, and the spectral energy
depends only on the modulus of the wavevector $F(\vec k) = F(k)$. Then
these two functions are related by the Hankel transform
\begin{equation}
F(k) = 2\pi \int_0^\infty C(l) J_0(kl)l dl \ ,
\label{hankel}
\end{equation}
where $J_0$ is the Bessel function and $E(k)=2\pi k F(k)$.

Another often used characteristic for scaling studies is the {\it
structure function\/} defined for arbitrary order $q$ as
\begin{equation}
S_q(l) =\  < (f(\vec x)-f(\vec x -\vec l))^q >_{|\vec l|=l}\ ,
\label{struc_q}
\end{equation}
where the brackets $<...>$ mean the average value. Calculation
of high-order structure functions requires a high accuracy of the
initial data. In the case of maps of external galaxies, where a
relatively small number of grid points is available and the noise is
significant, only the second-order function $S_2$, corresponding to the
energy spectrum (\ref{en_den}), can be discussed.

In the wavelet representation the scale distribution of the energy
can be characterized by the {\it wavelet spectrum}, defined as the energy
of the wavelet coefficients of scale $a$ of the whole physical plane
\begin{equation}
M(a) = \int_{-\infty}^{+\infty}
\int_{-\infty}^{+\infty} |W(a,\vec x)|^2 d \vec x \ .
\label{w_spec}
\end{equation}
The wavelet spectrum can be related to the Fourier spectrum. Using
(\ref{w_four}) one can easily rewrite (\ref{w_spec}) in the form
\begin{equation}
M(a) = {a^{4-2\kappa}\over{16\pi^4}}\int_{-\infty}^{+\infty}
\int_{-\infty}^{+\infty} |\hat f(\vec k)|^2
|\hat \psi ( a \vec k)|^2 d \vec k \ .
\label{w_spec_four}
\end{equation}
This relation shows that the wavelet spectrum is a smoothed version of
the Fourier spectrum. In the isotropic case (\ref{w_spec_four}) has a
more simple form
\begin{equation}
M(a) = {a^{4-2\kappa}\over{8\pi^3}}\int_{0}^{\infty}
E(k) |\hat \psi ( a k)|^2 d k \ .
\label{w_spec_four_iso}
\end{equation}

A generic property of fully developed small-scale turbulence is the
presence of an inertial range of scales in which the energetic
characteristics $E(k)$, $S_2(l)$ and $M(a)$ follow power laws. Let us
consider the relation between the spectral indices of these
characteristics. Let the structure function follow a power law
\begin{equation}
S_2(l) \sim l^\lambda \ .
\label{exp1}
\end{equation}
Then, $F(k) \sim k^{-\lambda-2}$ and the spectral energy density
exhibits a power law
\begin{equation}
E(k) \sim k^{-(\lambda+1)} \ .
\label{exp2}
\end{equation}
The behaviour of the wavelet spectrum M(a) depends on the normalization
$\kappa$ in definition (\ref{w_trans}). We will use $\kappa =2$,
which gives the same power law for the wavelet spectrum as for the
structure function
\begin{equation}
M(a) \sim a^{\lambda} \ .
\label{exp3}
\end{equation}
Choosing $\kappa=3/2$ the wavelet spectrum becomes $M(a) \sim
a^{\lambda+1}$, which is convenient if the result should be compared
with the Fourier spectrum $E(k)$. It should be noted that the scale
parameter $a$ commonly used in wavelets has the same meaning as
the distance $l$ in the autocorrelation (\ref{autocor}) or structure
function (\ref{struc_q}). Below we shall normally use the character $a$
for the scale parameter in any spectral characteristic.

An important remark concerns the power-law search in spectra. Every
wavelet has its own spectral portrait and the tail of the corresponding
spectral energy distribution often itself follows a power law. The
power index of this law defines the absolute limit of the spectral
slope that can be obtained using the given wavelet.

Note that the traditional calculation of the structure function $S_2$
following (\ref{struc_q}) can be interpreted as the calculation of
the wavelet spectrum (\ref{w_spec}), obtained using a ``special''
anisotropic wavelet, which gives the difference between two delta
functions separated by the unit distance
\begin{equation}
\psi(\vec x,\vec e) = \delta(\vec x)-\delta(\vec x - \vec e) \ ,
\label{special_w}
\end{equation}
where $\vec e$ is the unit vector. This quasi-wavelet (\ref{special_w})
is a bad wavelet in the sense that being very well localized in
physical space it has inevitably a very bad spectral resolution. This
means that structure functions give a poor scale separation and as a
result they are always very smooth. This is why the structure functions
can be useful only in the case of a well developed range of scales for
which a power law is established.

\begin{figure}
\centerline{\psfig{file=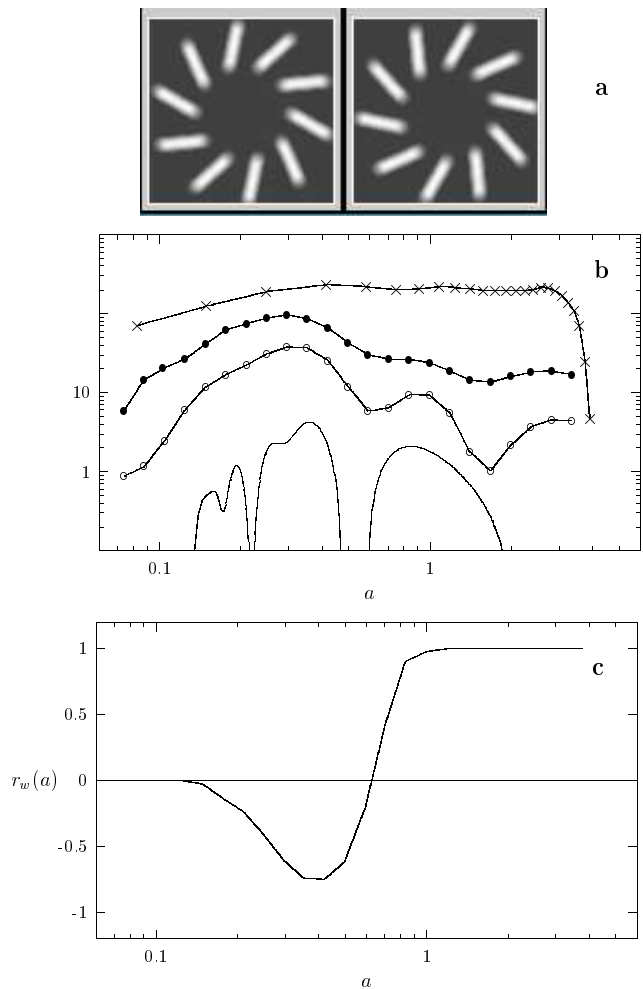,width=8.4truecm,%
       bbllx=137pt,bblly=149pt,bburx=491pt,bbury=699pt}}
\caption{Spectral and correlation characteristics of artificial images.
{\bf a} Test images with sides of length 2;
{\bf b} spectral characteristics of a single image:
second-order structure function (crosses), MH wavelet spectrum (black
dots), PH wavelet spectrum (circles), Fourier spectrum (line)
(the spectra are shifted along the vertical axis for better
presentation), $a=2\pi/k$;
{\bf c} wavelet cross-correlation coefficient.}
\label{r1}
\end{figure}

We conclude this section with an illustrative example which is
relevant to further data analysis. In Fig.~\ref{r1}a two images are
presented, the second one being a phase-shifted copy of the first one.
The spectral properties of these two images are identical and are shown
in Fig.~\ref{r1}b. The upper curve shows the structure function $S_2$
(\ref{struc_q}), which is almost flat over a large range of scales.
A steep decrease appears at large scales when the distance $l$
surpasses the total size of the structure. The next two curves
reproduce the wavelet spectra obtained with different wavelets,
MH (\ref{hat}) and PH (\ref{myf}). The graphs clearly illustrate the
increase of spectral resolution: some maxima are indicated in the MH
spectrum and three maxima are well pronounced in the PH spectrum.
These maxima
correspond to the scale of the width of an individual spoke in the
image, the scale of its length and the global scale of the structure.
The last (lower) curve shows the Fourier spectrum, calculated via the
Hankel transform (\ref{hankel}) of the autocorrelation function
(\ref{autocor}). It displays of course the best spectral resolution
but includes the upper harmonics -- two maxima at $a<0.25$ are the
double harmonics of the central structure. Note that the Fourier
spectrum is shown versus the physical scale $a$ which can be
interpreted as the wavelength, $a=2\pi /k$, thus the high frequencies
are to the left (small $a$).

This example illustrates the usual problem of Fourier presentation
of observational data -- real data give very spiky spectra and the
interpretation of different maxima in them is difficult. A reliable
choice of wavelets leads to smooth spectra which nevertheless clearly
indicate the dominant scales in the analysed data. Note also that the
global scale indicated by the structure function and the wavelets at
$a \approx 3$ is not visible in the Fourier spectrum -- the size of the
image is too small to contain the corresponding Fourier harmonic.

\section{Wavelet cross-correlations}

In this section we discuss correlations of images of the same
extragalactic object at different wavelengths and how they can be
quantified.

Let us consider two images (maps), $f_1(x,y)$ and $f_2(x,y)$,
with the same angular resolution and the same number of pixels.
The simplest way for a linear correlation study is the direct
calculation of the correlation coefficient point by point (pixel by
pixel)
\begin{equation}
r_p = {{\sum (f_{1i} - <f_1>)(f_{2i} - <f_2>)} \over
{((f_{1i} - <f_1>)^2(f_{2i} - <f_2>)^2)^{1/2}}} \ .
\label{cor_p}
\end{equation}
The accuracy of this estimate depends on the degree of correlation
and on the number of {\it independent\/} points $n$ (see for example
Edwards (1979))
\begin{equation}
\Delta r = {{\sqrt{1-r^2}} \over {\sqrt{n-2}}} \ .
\label{exp4}
\end{equation}
Here $n$ is the ratio of the map area to the area of the beam
($n \simeq 1700$ in the maps discussed in Section~5).

The ``pixel by pixel'' correlation coefficient is a global
characteristic which contains all the scales present in the
images, including the largest one which reflects the fact that the
centre of the extended intensity distribution is nearly at the same
position at most wavelengths (see the last panel of
Fig.~\ref{red_slices}).

It is interesting to investigate whether the correlation coefficient
depends on scale. Therefore we introduce the correlation coefficient
for a given scale $a$
\begin{equation}
r_w(a) = {{\int \int W_1(a,\vec x) W_2^*(a,\vec x) d \vec x} \over
{(M_1(a)M_2(a))^{1/2}}} \ .
\label{cor_w}
\end{equation}

To estimate the error $\Delta r_w(a)$ we use formula (\ref{exp4})
taking $n = (L/a)^2$, where $L$ is the size of the map.

To illustrate the behaviour of this wavelet correlation (i.e.
correlation on a given scale)
\footnote{The wavelet correlation was first introduced by
Nesme-Ribes et al. (1995) for two time series of solar data.}
let us return to the example given in
Fig.~\ref{r1}. The two images shown in the upper panel have an angular
shift $\pi/10$ (the spokes in the second map are just in between those
in the first one). Calculating the ``pixel by pixel'' correlation
coefficient yields $r_p= -0.01$. This vanishing value of the
correlation coefficient reflects the fact that the bright details
of both images occupy a small part of the image and do not overlap.
At the same time it is clear that the images are very well correlated
in general, because the images present two identical structures, one
of which is rotated.
In Fig.~\ref{r1}c we present the
wavelet correlation coefficient $r_w(a)$ calculated for these two
images. The plot explicitly shows that only the smallest scales are
really non-correlated. On the scale corresponding to the mean
distance between the spokes a strong anticorrelation is visible
($r_w \simeq -0.75$). Note that the minimum in $r_w$ does not coincide
with the maximum in the spectra -- the latter corresponds to the width
of an individual spoke, and the former to the distance between them. On
the scales which characterize the length of the spokes, the images
become correlated and $r_w$ tends to unity. On larger scales the
pictures are identical.

The relationship between the correlation coefficient $r_p$ and the
wavelet correlation $r_w$ can be obtained using (\ref{w_four}),
(\ref{w_spec}), (\ref{cor_w}), and in the case under consideration
($\kappa=2$) has the form
\begin{equation}
r_p = {{\int r_w(a) (M_1(a) M_2(a))^{1/2} a^{-1} da} \over
{(\int M_1(a) a^{-1} da \int M_2(a) a^{-1} da)^{1/2}}} \ .
\label{rp_rw}
\end{equation}

We conclude this section by a comparison of the introduced wavelet
correlation with the  cross-correlation function, defined as
\begin{equation}
h(\vec l) = \int_{-\infty}^{+\infty} \int_{-\infty}^{+\infty}
f_1(\vec x)f_2(\vec l - \vec x) d\vec x \ .
\label{crosscor}
\end{equation}
Using (\ref{w_four}),(\ref{hinchin}),(\ref{w_spec_four}) and
(\ref{cor_w}) one gets the following relation
\begin{equation}
r_w(a) = {{\int \int \hat h(\vec k)  |\hat \psi ( a \vec k)|^2
d \vec k } \over {(\int \int
\hat C_1(\vec k)  |\hat \psi ( a \vec k)|^2 d \vec k
\int \int \hat C_2(\vec k)  |\hat \psi ( a \vec k)|^2 d
\vec k)^{1/2}}} \ ,
\label{cor_ww}
\end{equation}
which in the limit of a wavelet with ideal scale resolution (wavelet
tends to become a harmonic) leads to a relation which in the isotropic
case can be written as
\begin{equation}
r_w(a) = {{\hat h(2\pi /a)} \over {(\hat C_1(2\pi /a)\hat
C_2(2\pi /a))^{1/2}}} \ .
\label{cor_w_trad}
\end{equation}

We emphasize that all the relations between wavelet coefficients,
Fourier decomposition and correlation functions are exact only if the
limits of the integrals surpass the real range of scales (spatial
frequencies) present in the object. From the practical point of view,
in the case of noisy data with relatively poor spatial resolution, the
calculation of $r_w(a)$ via wavelets has a strong advantage because in
this way we first separate the structures of a given scale present in
the data. The correct calculation of the Fourier transform $\hat
h(2\pi /a)$, however, implies accurate knowledge of the
cross-correlation function (\ref{crosscor}) in the whole range of
scales (the same is true for the spectral energy density and the
autocorrelation function).

In the simple case of the test in Fig.~\ref{r1}, both wavelets (MH and
PH) give practically the same correlation function $r_w(a)$. In this
purely symmetric example the same result can be obtained using
Eq.~(\ref{cor_w_trad}). The Fourier decomposition was used by
Elmegreen et al. (1992) to search for simple symmetries in galactic
images. However, the symmetry in such maps is generally weak and
superimposed onto a noisy background. Reliable statistics will then be
important to get significant results. Therefore we prefer the wavelet
with the best spatial localization and we use the MH wavelet for the
correlation study which occupies a smaller area than the PH wavelet for
the same scale. We note again that every Fourier harmonic covers the
whole plane so that its spatial location is completely absent.

\section{Analysis of scales and structures in optical, radio and infrared
maps of NGC\,6946}

\subsection{The data}

\begin{figure*}
\hbox to \hsize{
\psfig{file=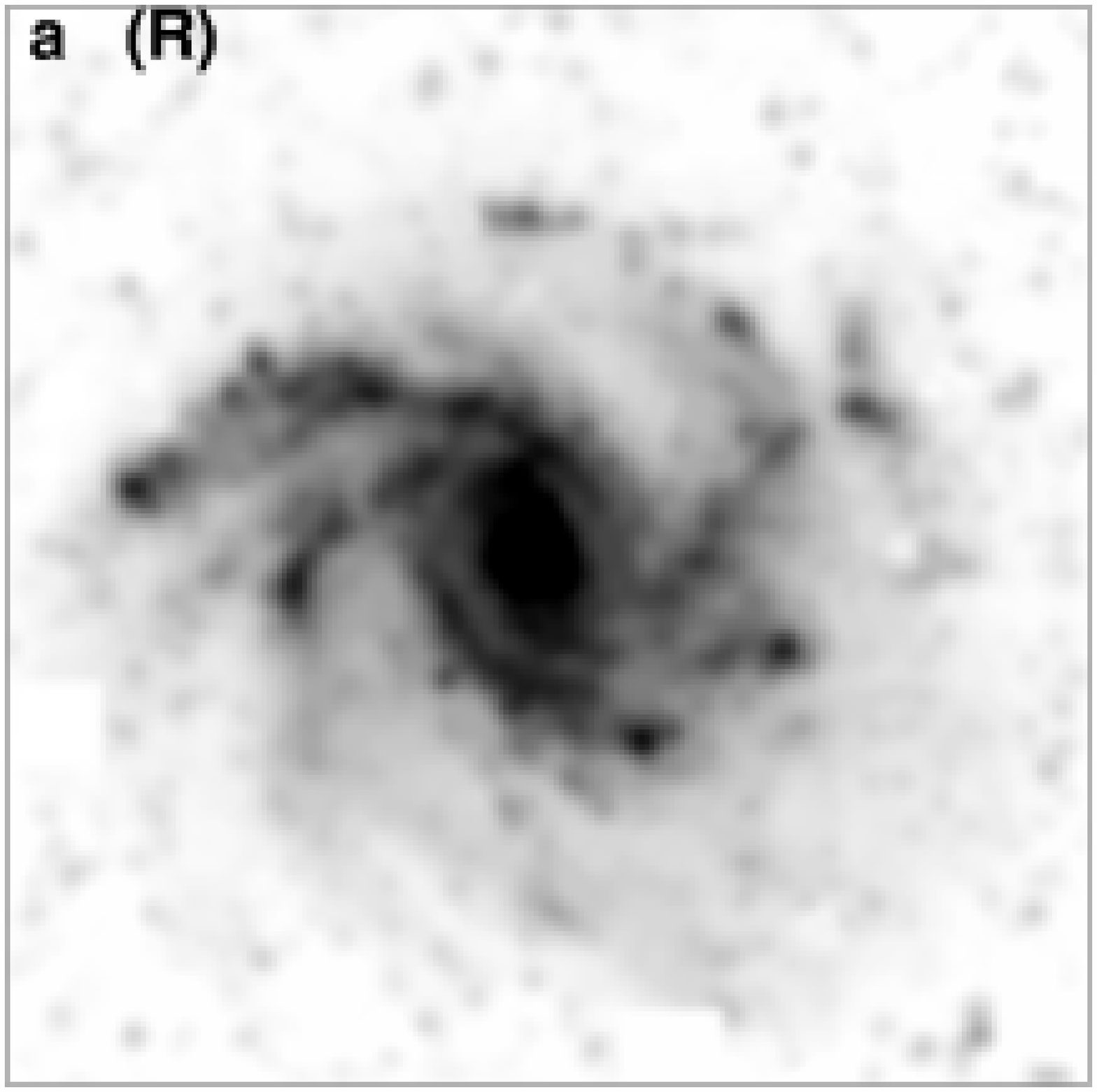,angle=270,width=5.6truecm,%
       bbllx=14pt,bblly=14pt,bburx=582pt,bbury=560pt}
\hspace{0.1truecm}
\psfig{file=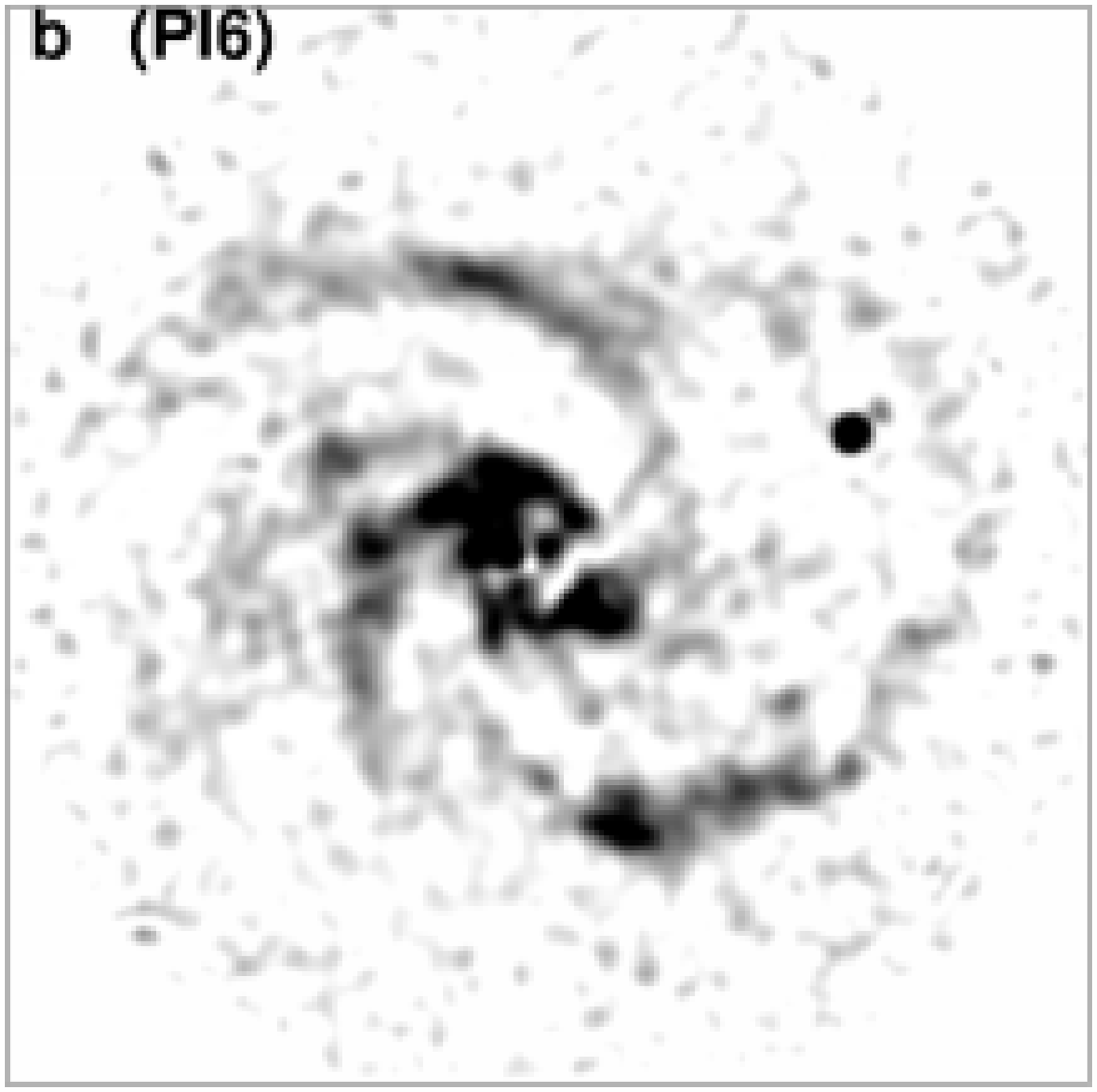,angle=270,width=5.6truecm,%
        bbllx=14pt,bblly=14pt,bburx=582pt,bbury=560pt}
\hspace{0.1truecm}
\psfig{file=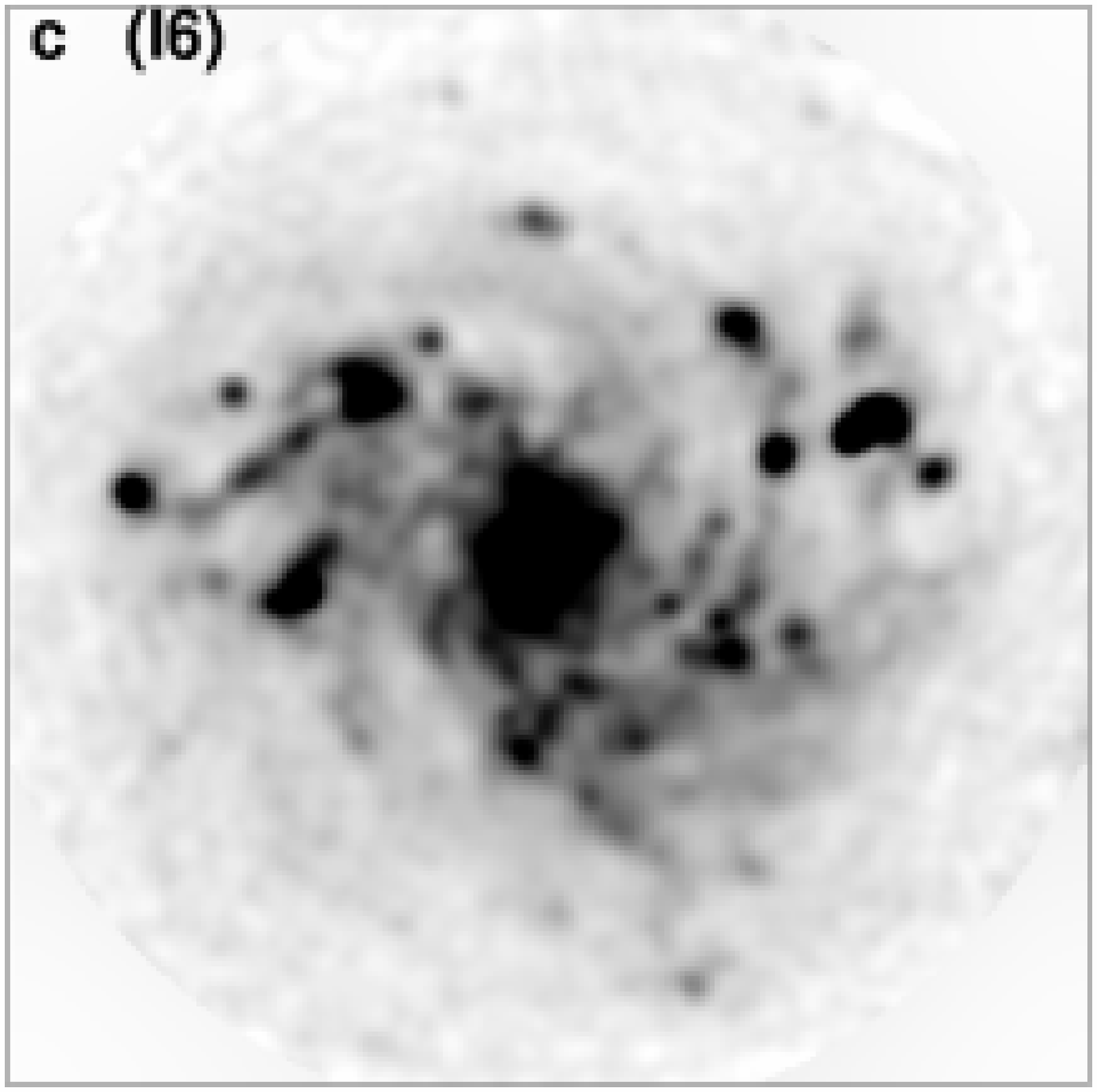,angle=270,width=5.6truecm,%
        bbllx=14pt,bblly=14pt,bburx=582pt,bbury=560pt}
}
\hbox to \hsize{
\psfig{file=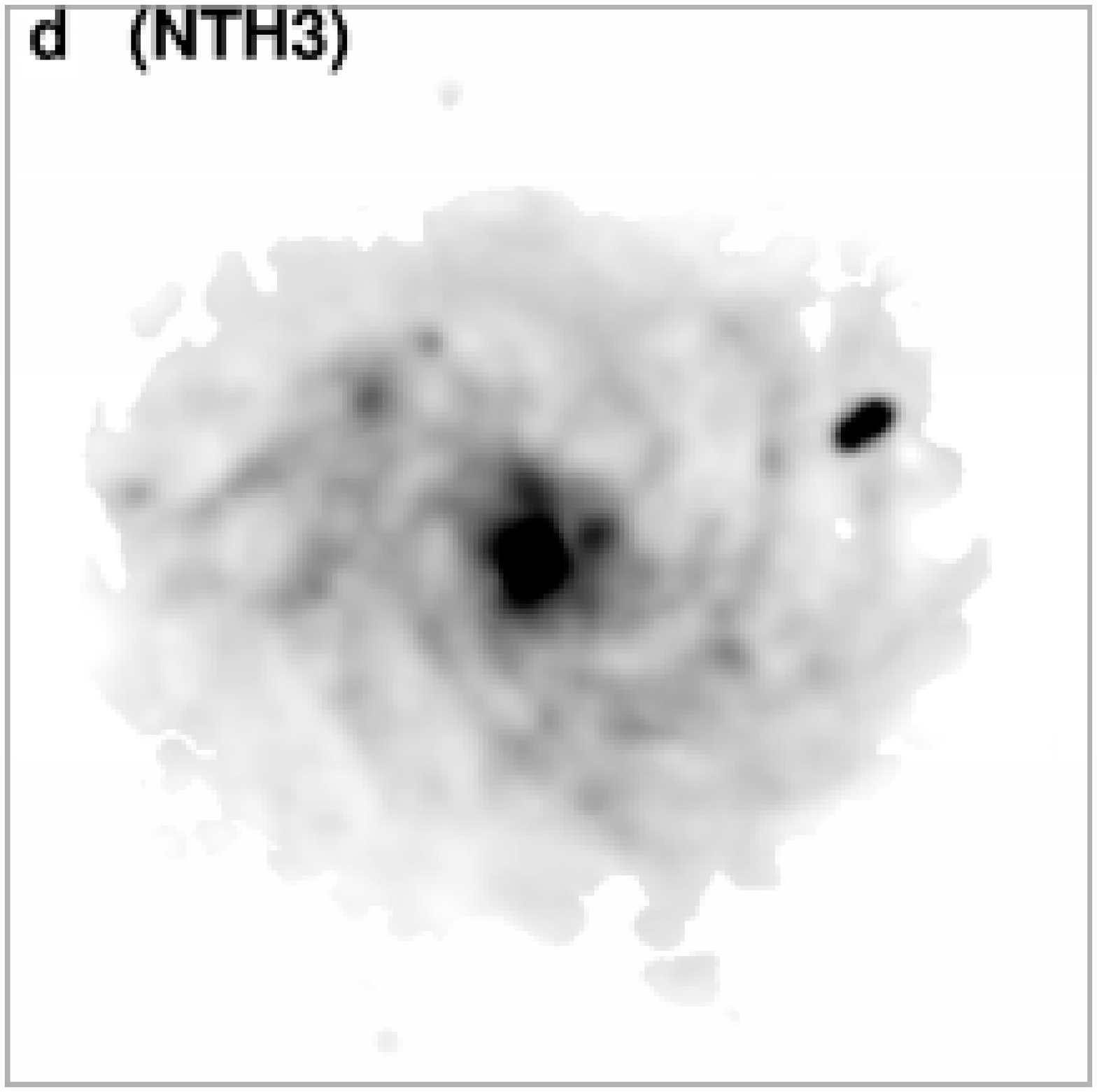,angle=270,width=5.5truecm,%
        bbllx=14pt,bblly=14pt,bburx=582pt,bbury=560pt}
\hspace{0.2truecm}
\psfig{file=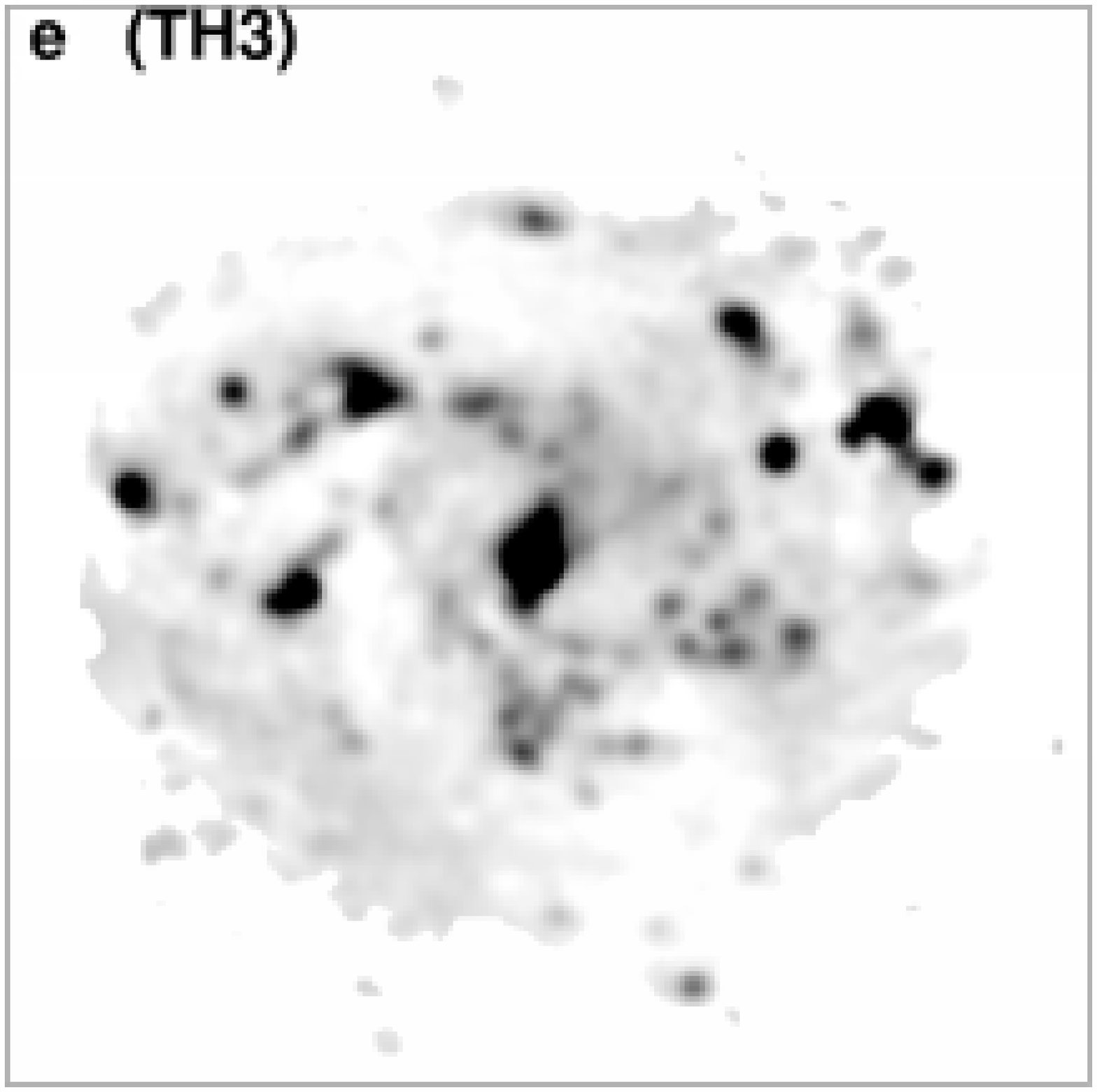,angle=270,width=5.5truecm,%
        bbllx=14pt,bblly=14pt,bburx=582pt,bbury=560pt}
\hspace{0.2truecm}
\psfig{file=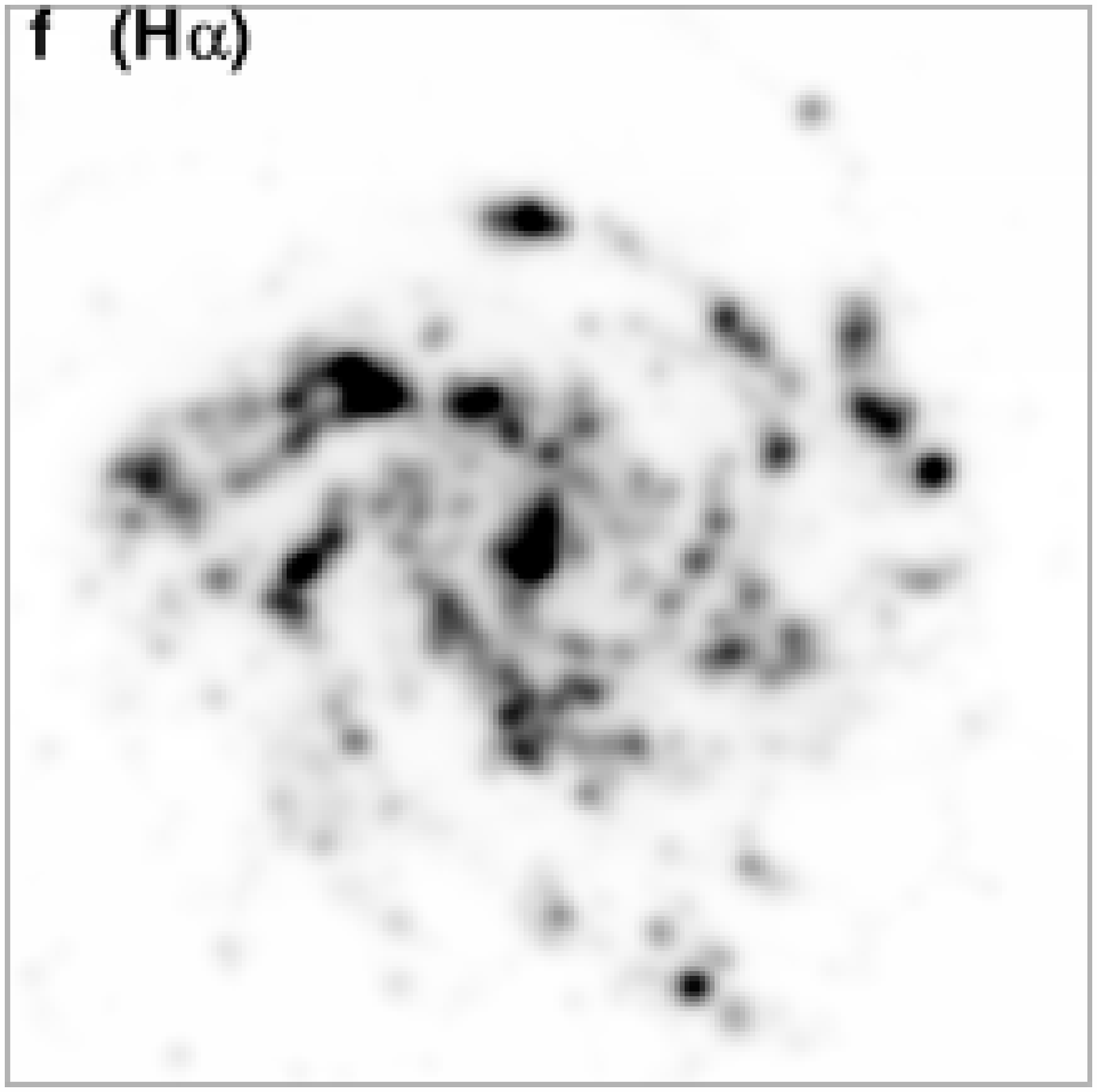,angle=270,width=5.5truecm,%
        bbllx=14pt,bblly=14pt,bburx=582pt,bbury=560pt}
}
\hbox to \hsize{
\psfig{file=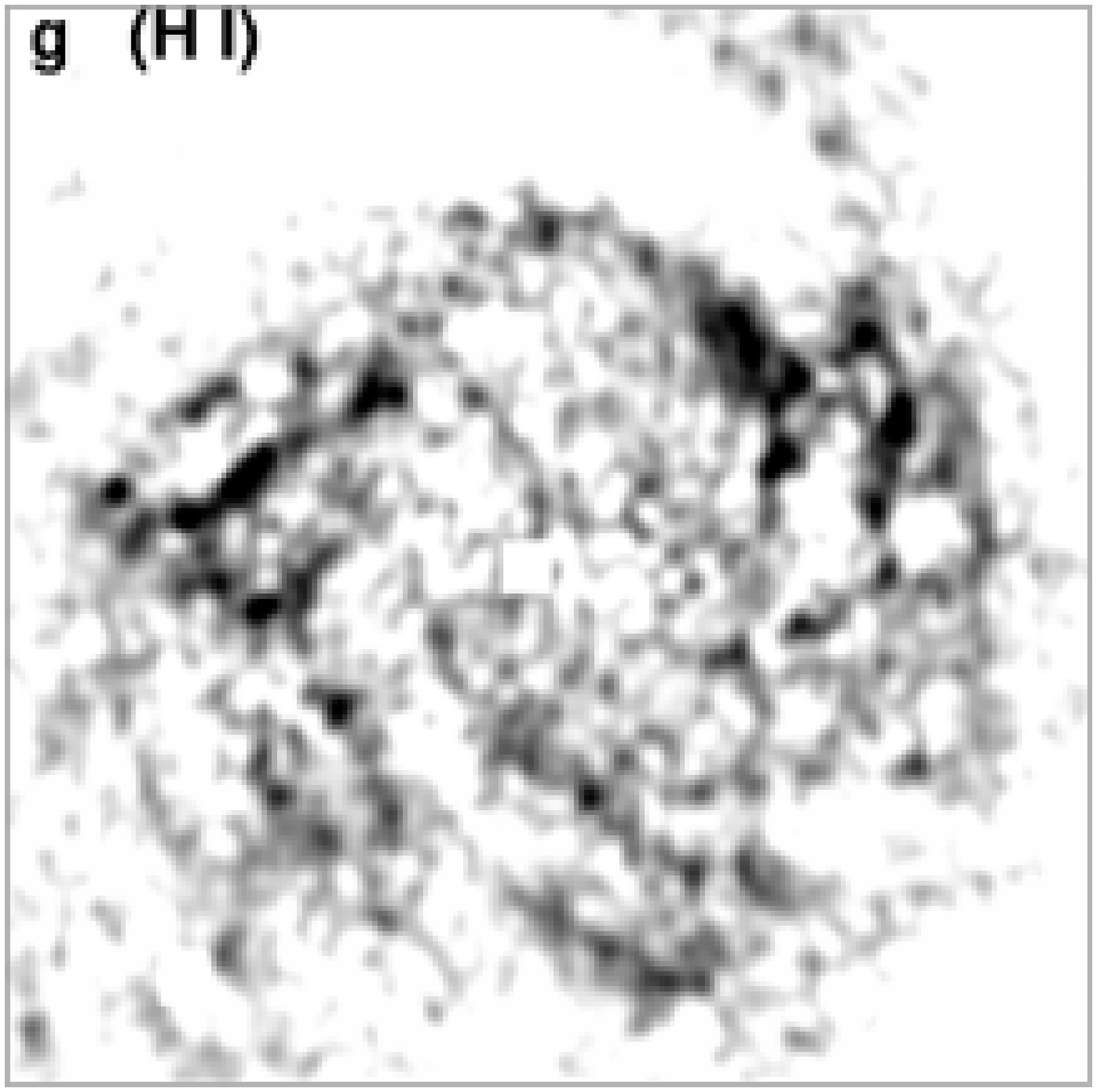,angle=270,width=5.5truecm,%
        bbllx=14pt,bblly=14pt,bburx=582pt,bbury=560pt}
\hspace{0.2truecm}
\psfig{file=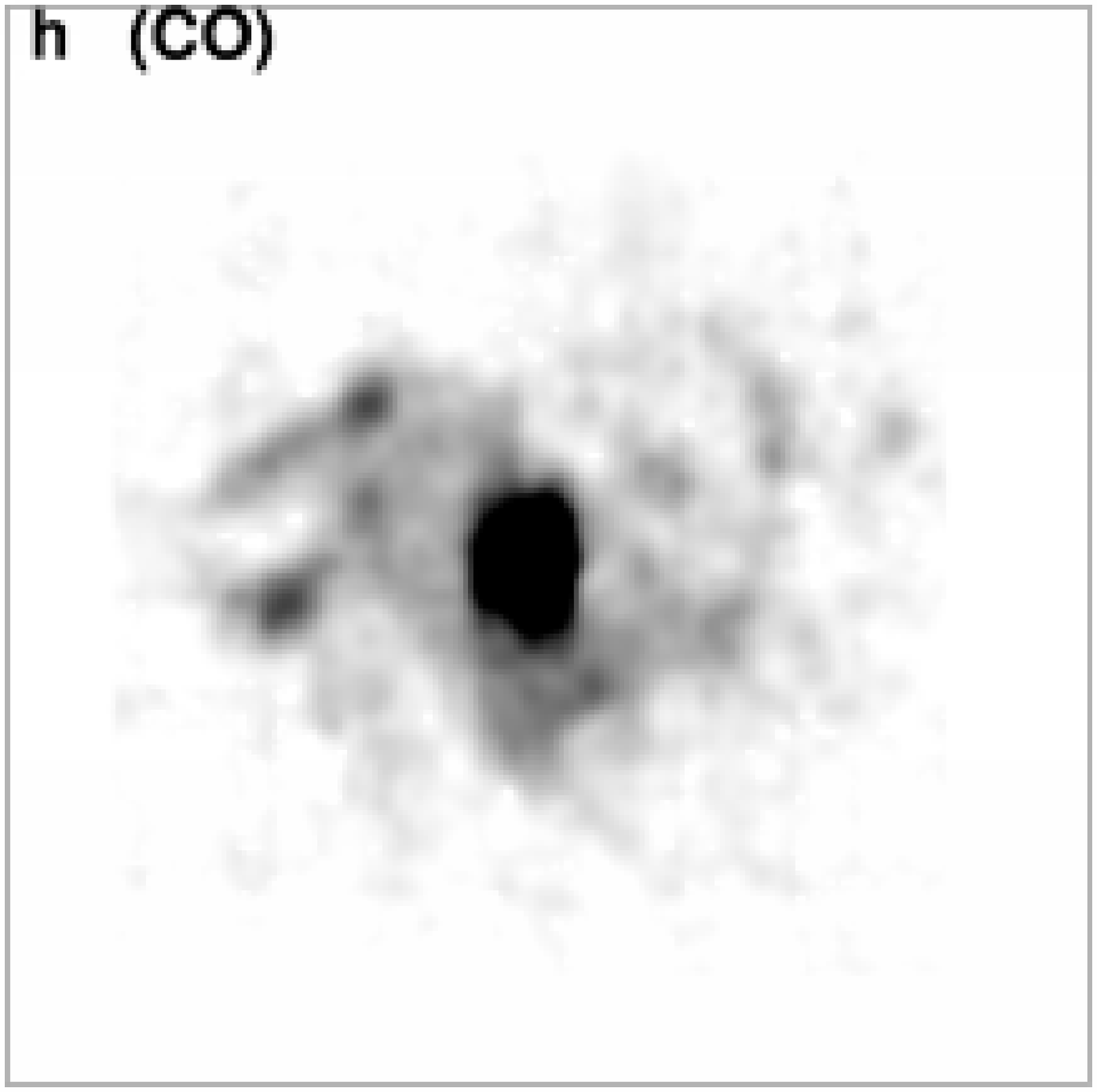,angle=270,width=5.5truecm,%
        bbllx=14pt,bblly=14pt,bburx=582pt,bbury=560pt}
\hspace{0.2truecm}
\psfig{file=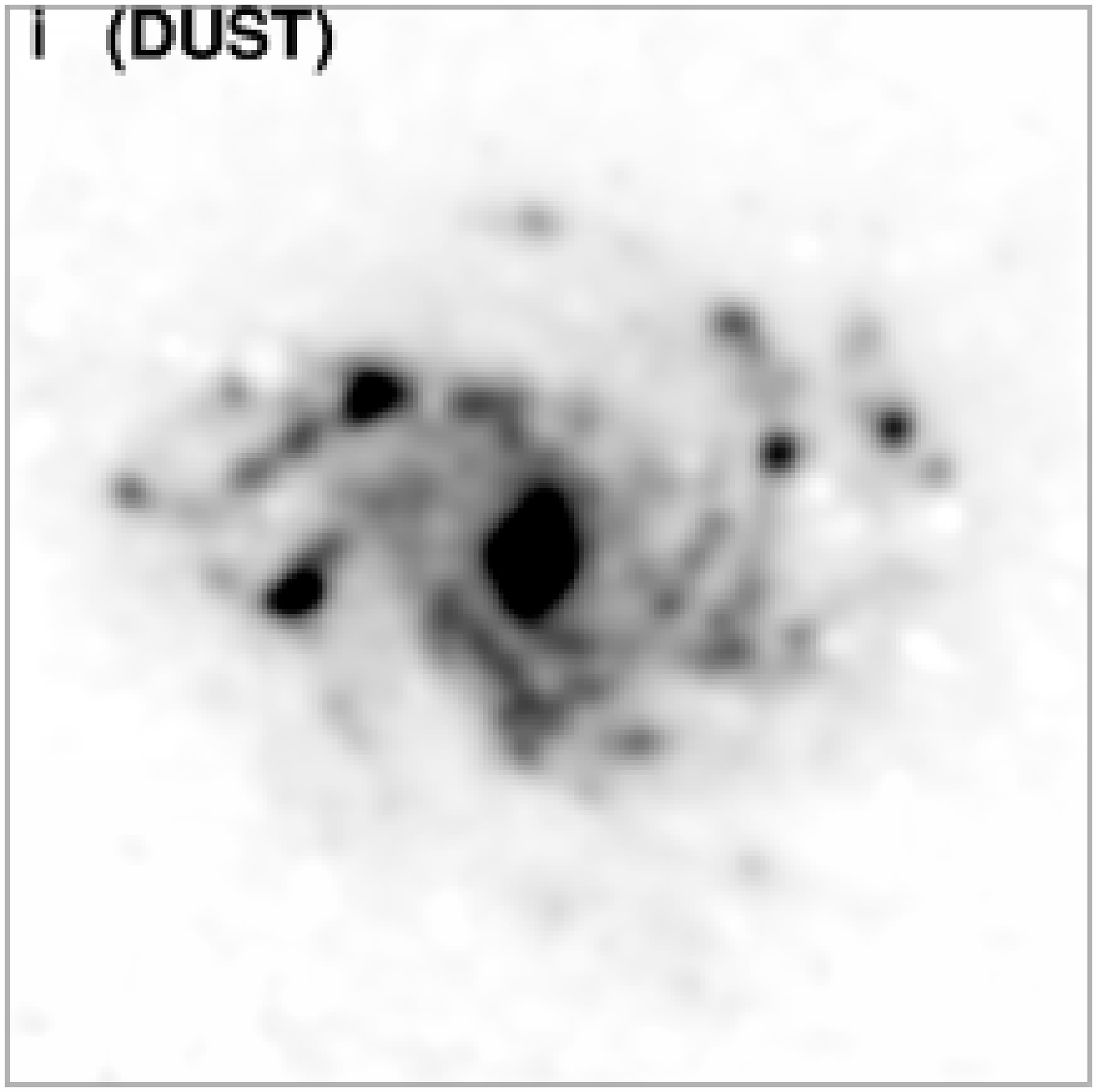,angle=270,width=5.5truecm,%
        bbllx=14pt,bblly=14pt,bburx=582pt,bbury=560pt}
}
\caption{Images of NGC\,6946 in different spectral ranges.
Top row (from left to right):
{\bf a} optical (red) light,
{\bf b} linearly polarised radio continuum emission at $\lambda$6.2~cm,
{\bf c} total radio continuum emission at $\lambda$6.2~cm.
Middle row:
{\bf d} nonthermal radio continuum emission at $\lambda$3.5~cm,
{\bf e} thermal radio continuum emission at $\lambda$3.5~cm,
{\bf f} H$\alpha$ line emission of ionised hydrogen $\HII$.
Bottom row:
{\bf g} line emission of neutral atomic hydrogen $\HI$,
{\bf h} (1-0) line emission of the CO molecule,
{\bf i} mid-infrared emission of warm dust and PAH particles at 12--18\,$\mu$m.
The size of every image is $11\arcmin \times 11\arcmin$ and the beam diameter
is always $15\arcsec$.}
\label{images}
\end{figure*}

NGC\,6946 is a nearby spiral galaxy for which a distance of 5.5~Mpc is
usually assumed. Data in many spectral ranges are available. For
our analysis we used:

a) The map of the optical broadband emission in red light (R) from
the digitized Palomar Sky Survey.\footnote{Based on photographic data
of the National Geographic Society -- Palomar Observatory Sky Survey
(NGS--POSS) obtained using the Oschin Telescope on Palomar Mountain.
The NGS--POSS was funded by a grant from the National Geographic
Society to the California Institute of Technology. The plates were
processed into the present compressed digital form with their
permission. The Digitized Sky Survey was produced at the Space
Telescope Science Institute under US Government grant NAG W--2166.}
We subtracted foreground stars from the map and smoothed it to the
angular resolution of the radio continuum maps. The result is shown in
Fig.~\ref{red_slices}.

b) The map of the linearly polarised radio continuum emission at
$\lambda$6.2~cm (PI6), combined from observations with the VLA synthesis
telescope operated by the NRAO\footnote{The National Radio Astronomy
Observatory is a facility of the National Science Foundation operated
under cooperative agreement by Associated Universities, Inc.} and the
Effelsberg single dish operated by the MPIfR (Beck \& Hoernes 1996). The
angular resolution is $15\arcsec$ which corresponds to 400~pc at the
assumed distance to NGC\,6946 (5.5~Mpc).

c) The map of the total radio continuum emission at $\lambda$6.2~cm
(I6), combined from observations with the VLA synthesis telescope and
the Effelsberg single dish (Beck \& Hoernes 1996). The angular
resolution is $15\arcsec$. We subtracted a strong unresolved source on
the northwestern edge of NGC\,6946 because it seems unrelated to the
galaxy.

d) The map of the nonthermal (synchrotron) radio continuum emission at
$\lambda$3.5~cm (NTH3), derived from the spectral index map between
$\lambda$3.5~cm and $\lambda$20.5~cm at $15\arcsec$ resolution
(Beck, in prep.) assuming a constant synchrotron spectral index of
$\alpha=-0.9$ (where the flux density $S$ varies with frequency $\nu$
as $S_{\nu} \propto \nu^{\alpha}$). The spectral index and nonthermal
intensity were computed only for pixels where both intensity values
exceed 10 times the rms noise.

e) The map of the thermal radio continuum emission (free-free emission
from ionised hydrogen) at $\lambda$3.5~cm (TH3) which is the difference
map between the total and the synchrotron emission.

f) The map of the H$\alpha$ line emission (H$\alpha$) of ionised
hydrogen $\HII$, integrated over the whole frequency width of the line
(Ferguson et al. 1998). We subtracted
foreground stars and smoothed the map to the angular resolution of the
radio continuum maps. The H$\alpha$ map
by Ferguson et al. (1998) has a higher signal-to-noise ratio
than the map used in Frick et al. (2000).

g) The map of the radio line emission of neutral atomic hydrogen $\HI$
at $\lambda$21.1~cm ($\HI$) observed with the Westerbork Synthesis Radio
Telescope, integrated over the whole frequency width of the line
(Kamphuis \& Sancisi 1993). The $\HI$ map has an original resolution of
$13\arcsec \times 16\arcsec$ and was not smoothed.

h) The map of the (1-0) line emission of the molecule CO at
$\lambda2.6$~mm (CO) observed with the IRAM 30-m single-dish telescope,
integrated over the whole frequency width of the line (Walsh et al.
2001). The CO map has an original resolution of $22\arcsec$ and was not
smoothed. The lower resolution of the CO map compared with the other
maps may influence the comparison on small scales, but the effect
turns out to be small (see Fig.~\ref{cor_red_pi}).

i) The map of the mid-infrared emission (12--18\,$\mu$m) (DUST) observed
with the ISOCAM camera (filter LW3) on board of the ISO satellite
(Dale et al. 2000). This wavelength range is dominated by the band of
emission lines from polycyclic aromatic hydrocarbons (PAHs) at
12.7\,$\mu$m, but also includes continuum emission from small, warm dust
grains. The LW3 map has an original resolution of about $8\arcsec$ and
was smoothed to a beamwidth of $15\arcsec$.

The map in LW2 (5--8.5\,$\mu$m) is fully dominated by PAH emission
bands. As it is very similar to the LW3 map, we did not include it in
our analysis.

All maps were transformed to the same area ($11\arcmin \times 11\arcmin$)
and pixel size ($2\arcsec$). Each map contains $331\times 331$ pixels.
Plots of these images are shown in Fig.~\ref{images}.

In the radio continuum maps, we removed the bright double source at the
western edge of the galaxy which has no counterpart in any other
spectral range and thus is most probably unrelated to NGC\,6946.
Finally, we removed the central region in 7 of the 9 maps
by subtracting fitted Gaussian sources with half-power widths of about
$0 \farcm 3$ (infrared, total, nonthermal and thermal radio),
$0 \farcm 5$ (H$\alpha$, CO) and
$1\arcmin$ (red light). This avoids artefacts produced by improper
continuum background subtraction in the central region of the H$\alpha$
map as well as contamination of the mid-infrared map by enhanced
continuum emission from warm dust grains.

\subsection{Spectral characteristics}

Before starting the presentation of spectral characteristics of
maps of an external galaxy we want to emphasize that one should
be very careful when interpreting such kind of results for two
reasons. Firstly, for the scales that are reliably resolved, the
maps cannot be treated as homogeneous and
isotropically turbulent (as is possible when analysing the Milky
Way turbulence from Galactic surveys (Minter \& Spangler 1996)).
One can expect to find the dominant scale in the map at a certain
wavelength (if it exists) and the scale below which the
field under consideration can be interpreted as homogeneous -- that
is one may separate the large-scale structures and the
small-scale turbulence. Secondly, one has to carefully separate
the spectral properties of the analysed image from the spectral
properties of the analysing function and from structures caused by
observational problems (regions without observations, bright
sources, instrumental noise etc).

\begin{figure}
\centerline{\psfig{file=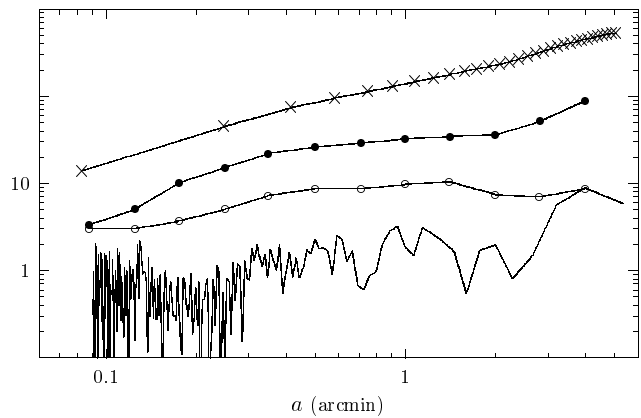,width=8.4truecm,%
       bbllx=111pt,bblly=314pt,bburx=431pt,bbury=523pt}}
\caption{NGC\,6946. Spectral characteristics of the map of total radio emission (I6):
second-order structure function (crosses), MH
wavelet spectrum (dots), PH wavelet spectrum (circles), Fourier
spectrum (line). The size of a pixel is $0\farcm 03$ and the beam
diameter is $0\farcm 25$.}
\label{sp_6IM}
\end{figure}

\begin{figure}
\centerline{\psfig{file=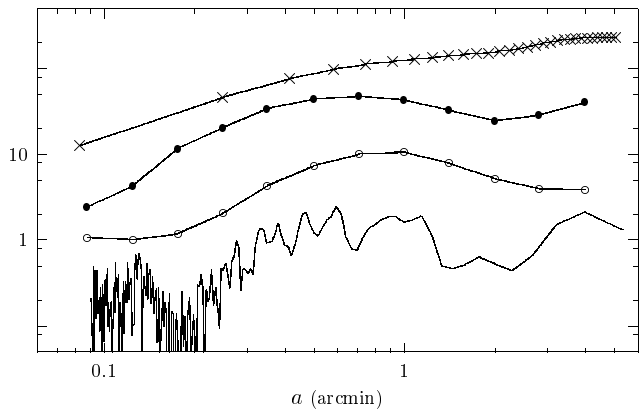,width=8.4truecm,%
       bbllx=107pt,bblly=314pt,bburx=431pt,bbury=523pt}}
\caption{NGC\,6946. Spectral characteristics of the H$\alpha$ map: second-order
structure function (crosses), MH wavelet spectrum (dots),
PH wavelet spectrum (circles), Fourier spectrum (line).
The size of a pixel is $0\farcm 03$ and the beam diameter
is $0\farcm 25$.}
\label{sp_HA}
\end{figure}

We first show the spectral characteristics as defined above for two
maps of NGC\,6946: the total radio emission (Fig.~\ref{sp_6IM})
and the distribution of the H$\alpha$ emission (Fig.~\ref{sp_HA}).
Clearly, the structure function totally smoothes the spectral energy
distribution suggesting a spectrum of fully developed turbulence on
all scales with a power law of slope 0.7. However, the other curves
show that such a spectrum is not present in the data.
The Fourier spectrum clearly displays the border
between the informative part of the spectrum and the small-scale noise,
visible as a flat (or even increasing, as in the H$\alpha$ spectrum)
part for scales smaller than 0\farcm 2. This scale is close to the size
of the beam (0\farcm 25) and proves the evident statement that little
can be said about scales below the scale of the beamwidth.
The structure function and the wavelet spectra show less
flattening on scales smaller than the beam.

Wavelet spectra again give an intermediate presentation of the spectral
properties. As the optimal spectral characteristic the PH wavelet
spectrum can be proposed which smoothes the numerous peaks in the
Fourier decomposition but is still sensitive enough to indicate the
scales containing most of the energy. The I6 spectrum  shows two
weak maxima at 0\farcm 5 and 1\farcm 3. Note that the spectrum obtained
by the MH wavelet does not separate these two maxima. The wavelet
spectrum of H$\alpha$ shows a well pronounced maximum at
0\farcm 8--1\farcm 0. Even this maximum is hardly visible in the
structure function.

\begin{figure}
\centerline{\psfig{file=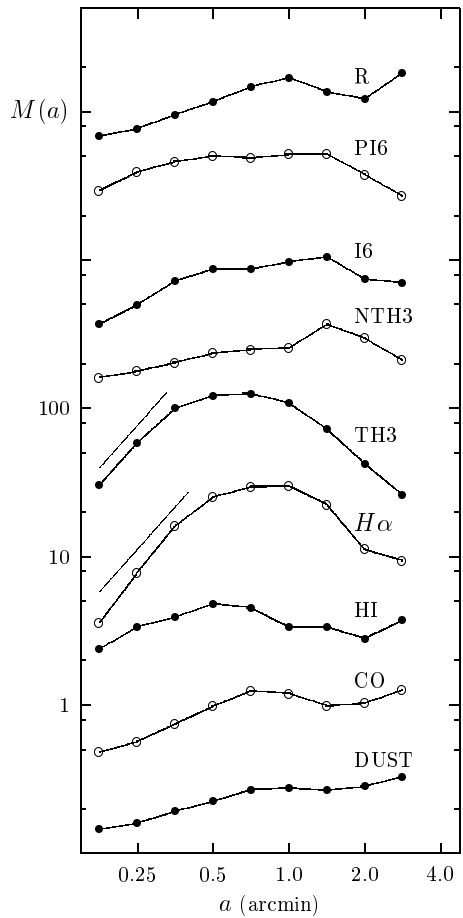,width=6.0truecm,%
       bbllx=103pt,bblly=206pt,bburx=313pt,bbury=631pt}}
\caption{NGC\,6946. Wavelet spectra (analysing wavelet is PH).
The thin lines above H$\alpha$ and TH3 indicate a slope of 5/3.}
\label{spectra_myf}
\end{figure}

In Fig.~\ref{spectra_myf} we present wavelet spectra calculated
for all 9 images of NGC\,6946 using the PH wavelet (\ref{myf}).
The range of scales is limited to $0\farcm 2 < a < 4\farcm 0$.
Comparing the spectral distributions we note that only H$\alpha$
and TH3 show a clear maximum (though shifted in scale, see
Sect.~6.1). Less strong maxima are visible in CO, $\HI$, R and
nonthermal emission NTH3. Almost flat are the spectra of total
radio emission I6 and polarised radio emission PI6 -- they both
show weak maxima near 0\farcm 5 and 1\farcm 2. The location of
the maxima differs for the images. The thermal emission TH3 and
$\HI$ contain most of the energy on the scale of about 0\farcm 5,
CO on 0\farcm 7, H$\alpha$ and R on 1\farcm 0, the NTH3 (and
maybe I6) on 1\farcm 4--1\farcm 5. For comparison: the typical
size of a giant complex of gas clouds is $\approx 1$~kpc (0\farcm
6), the typical width of the spiral arms is $\approx 1.5$~kpc
(0\farcm 9).

The spectra in Fig.~\ref{spectra_myf} show that the spatial resolution of
the available images is insufficient for the study of properties of the
small-scale turbulence. The inertial range is expected to start at
scales several times smaller than the dominant scale (turbulent
macroscale). If the observed object contains the classical
three-dimensional Kolmogorov turbulence with the spectrum $E(k)\sim
k^{-5/3}$ (corresponding to the structure function $S_2 \sim
l^{2/3}$), the two-dimensional projection should be characterized be a
structure function $S_2 \sim l^{5/3}$ (Spangler 1991). The spectral
slope of the projected field steepens because the smaller scales are
smoothed in the projection. In Fig.~\ref{spectra_myf}, only TH3 and
H$\alpha$
show Kolmogorov-type behaviour on small scales (see Sect.~6.1).

\begin{figure}
\centerline{\psfig{file=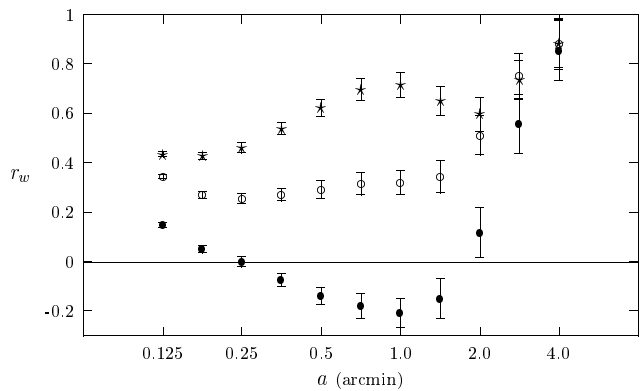,width=8.4truecm,%
       bbllx=86pt,bblly=314pt,bburx=431pt,bbury=527pt}}
\caption{NGC\,6946. Scale-by-scale wavelet cross-correlations:
H$\alpha$--PI6 (dots); H$\alpha$--TH3 (stars); H$\alpha$--NTH3
(circles).}
\label{cor_ha}
\end{figure}

\subsection{Cross-correlations in the galaxy NGC\,6946}

We start by calculating the classical ``pixel by pixel'' cross-correlation
coefficient $r_p$ for every pair of galactic maps under discussion. The
values of $r_p$ are given in the upper part (above the diagonal) of
Table~1. Note that for most pairs the value of the coefficient indicates
a statistically significant correlation. Only the PI6 map in pair with
TH3, H$\alpha$ and $\HI$ yields correlation coefficients significantly
less than 0.5 (0.23, 0.25 and 0.29, respectively).

\begin{table*}
\caption{Cross-correlations between 9 maps of NGC\,6946. The classical
correlation coefficients $r_p$ are given above the diagonal of the
table. Below the diagonal the mean values of the wavelet correlation
coefficient $r_w(a)$, calculated for the range of scales
$0\farcm 4 < a < 1\farcm 2$, are shown.}
\begin{tabular}{|l|c|c|c|c|c|c|c|c|c|}
\hline
& & & & & & & & & \\
       & R  & PI6  &  I6  &  NTH3  & TH3  &  H$\alpha$  & $\HI$ & CO & DUST
\\
& & & & & & & & \\  \hline
R    &  & $0.51$ & $0.84$ & $0.86$ & $0.66$ & $0.77$ & $0.44$ & $0.83$ &
$0.88$
\\
&   & $\pm 0.03$& $\pm0.02$& $\pm 0.02$& $\pm 0.03$& $\pm 0.02$& $\pm 0.03$&
$\pm 0.02$& $\pm0.02$\\ \hline
PI6 & $-0.11$ &   & $0.56$ & $0.67$ & $0.23$ & $0.25$ & $0.29$ & $0.53$ &
$0.49$
\\
& $\pm 0.06$ &  & $\pm 0.03$& $\pm 0.03$& $\pm 0.03$& $\pm 0.03$& $\pm
0.03$&
$\pm 0.03$& $\pm 0.03$\\ \hline
I6 & $0.49$ & $0.19$ &   & $0.92$ & $0.81$ & $0.77$ & $0.47$ & $0.78$ &
${\mathbf 0.94}$
\\
& $\pm 0.05$& $\pm 0.06$&   & $\pm 0.02$& $\pm 0.02$& $\pm 0.02$& $\pm
0.03$&
$\pm 0.02$& ${\mathbf \pm 0.02}$\\ \hline
NTH3 & $0.29$  &$0.36$ & $0.71$ &   & $0.60$ & $0.64$ & $0.40$ & $0.80$ &
$0.89$
\\
& $\pm 0.06$& $\pm 0.06$& $\pm 0.04$&  & $\pm 0.03$& $\pm 0.03$& $\pm 0.03$&
$\pm 0.02$& $\pm 0.02$\\ \hline
TH3 & $0.57$  & $-0.10$  & $0.77$ & $0.53$  &   & $0.81$ & $0.45$ & $0.57$ &
$0.79$\\
& $\pm 0.05$& $\pm 0.06$& $\pm 0.04$& $\pm 0.05$&  & $\pm 0.02$& $\pm 0.03$&
$\pm 0.03$& $\pm 0.02$\\ \hline
H$\alpha$& $0.68$ & $-0.22$ & $0.65$ & $0.34$  & $0.80$  &   & $0.50$ &
$0.65$ &
$0.81$ \\
& $\pm 0.05$& $\pm 0.06$& $\pm 0.05$& $\pm 0.06$& $\pm 0.03$&   & $\pm
0.03$&
$\pm 0.03$& $\pm 0.02$\\ \hline
$\HI$   & $0.35$ & $0.14$ & $0.47$ & $0.39$ & $0.37$  & $0.41$ &  & $0.41$ &
$0.43$ \\
& $\pm 0.06$& $\pm 0.06$& $\pm 0.06$& $\pm 0.06$& $\pm 0.06$& $\pm 0.06$&  &
$\pm 0.03$& $\pm 0.03$\\ \hline
CO     & $0.35$ & $0.24$ & $0.48$ & $0.32$ & $0.36$ & $0.36$ & $0.49$ &  &
$0.85$  \\
& $\pm 0.06$& $\pm 0.06$& $\pm 0.06$& $\pm 0.06$& $\pm 0.06$& $\pm 0.06$&
$\pm
0.06$&  & $\pm 0.02$\\ \hline
DUST & $0.56$ & $0.14$ & ${\mathbf 0.84}$ & $0.47$ & $0.63$ & $0.60$ & $0.57$ & $0.66$
&
 \\
& $\pm 0.05$& $\pm 0.06$& ${\mathbf \pm 0.04}$& $\pm 0.06$& $\pm 0.05$& $\pm 0.05$&
$\pm
0.05$& $\pm 0.05$&   \\ \hline
\end{tabular}
\end{table*}

For each pair of maps the wavelet correlation coefficient $r_w$ was also
calculated, using the MH wavelet. As an example, the behaviour of
$r_w(a)$ is shown in Fig.~\ref{cor_ha} for three pairs of data:
H$\alpha$--PI6, H$\alpha$--TH3 and H$\alpha$--NTH3.
The high correlation on largest scales reflects only the coincidence
of the extended regions corresponding to the galaxy's disk
(see Fig.~\ref{red_slices}, bottom right).
However, on intermediate scales the three curves
display very different behaviours. The {\it anticorrelation} between
H$\alpha$ and PI6 is of special interest (see Sect.~6.2).
The tendency of all curves in Fig.~\ref{cor_ha} to grow towards the smallest
scale ($a=0\farcm 125$) indicates that this scale is affected by
fluctuations on scales significantly smaller than the resolution of the maps
($0\farcm 25$) and thus will not be considered further, whereas
the next computed scale ($a=0\farcm 2$), though still slightly smaller than
the beam size, gives useful information.
Note that the number of independent points decreases towards larger scales
which results in increasing error bars on these scales.

\begin{figure*}
\centerline{\psfig{file=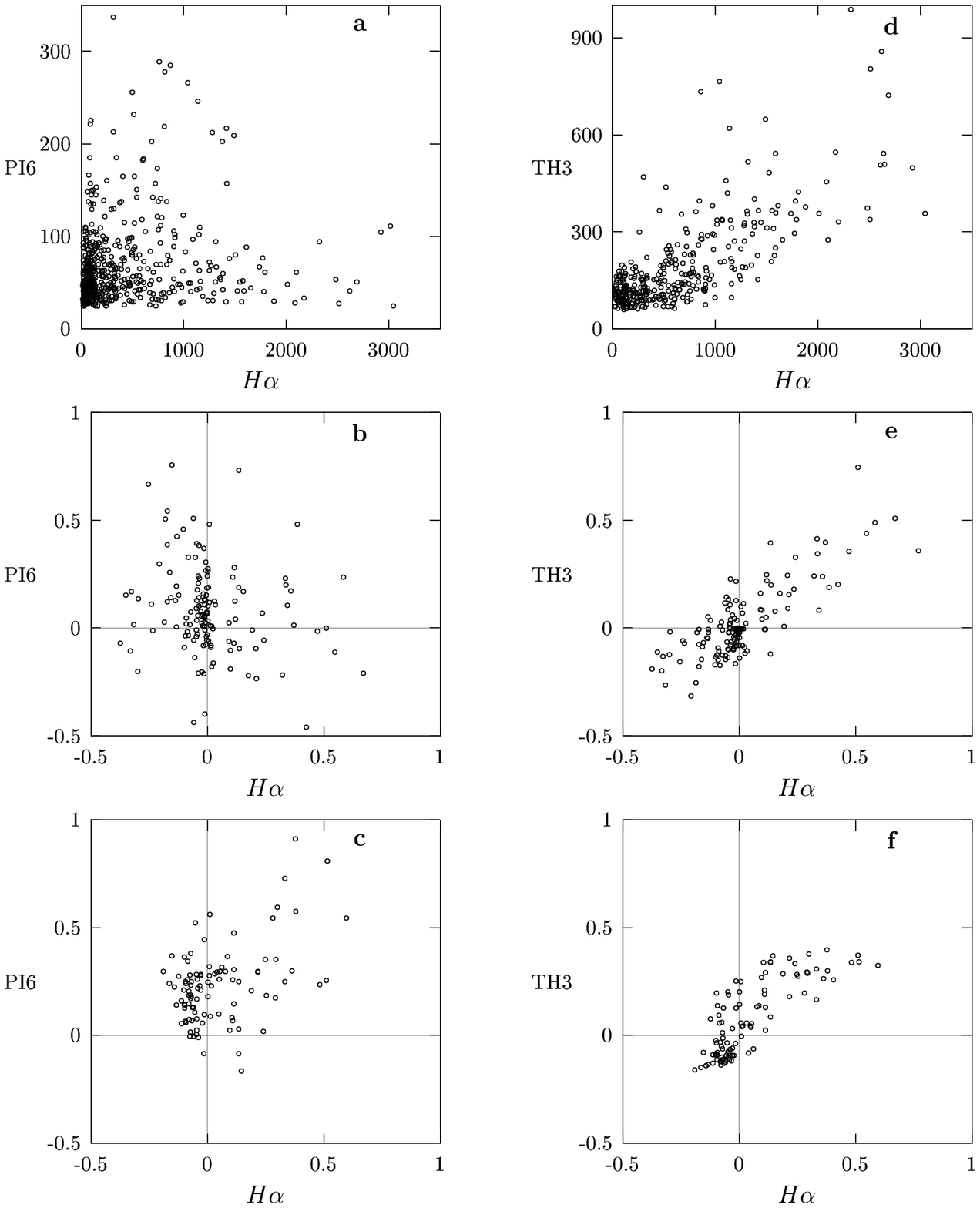,width=14truecm,%
       bbllx=84pt,bblly=130pt,bburx=552pt,bbury=707pt}}
\caption{NGC\,6946. Cross-correlations of the intensities of the initial images and
of their wavelet decompositions. Polarised radio emission (PI6) versus H$\alpha$:
{\bf a} initial data,
{\bf b} wavelet coefficients $w$ for scales $a=1\arcmin$,
{\bf c} for $a=2\farcm 8$.
Thermal radio emission (TH3) versus H$\alpha$:
{\bf d} initial data,
{\bf e} wavelet coefficients $w$ for scales $a=1\arcmin$,
{\bf f} for $a=2\farcm 8$.}
\label{cor_elly}
\end{figure*}

To illustrate the difference in correlating the initial data and
the wavelet decomposition on various scales, we show the corresponding
plots in Fig.~\ref{cor_elly} for two extreme cases of
Fig.~\ref{cor_ha}. The correlation of the maps PI6 and H$\alpha$
gives $r_p=0.25$. This low value of $r_p$ is understandable when looking
at the distribution of points in Fig.~\ref{cor_elly}a. In panel (b) we
show the wavelet coefficients $W(a,\vec x)$ obtained for the PI6 map
versus the corresponding wavelet coefficients calculated for the
H$\alpha$ map. The scale parameter is fixed at the value $a=1\farcm 0$,
which corresponds to the minimum in the curve $r_w(a)$ for this pair of
maps in Fig.~\ref{cor_ha}. The cloud of points becomes elongated along
the line $y = -x$ which corresponds to a negative value of $r_w$. Note
that the wavelet coefficients $W(a,\vec x)$ become negative in case of
local minima on a certain scale in the analysed data. Panel (c) shows
the corresponding plot for the same pair of maps, but for the scale
$a=2\farcm 8$, which gives $r_w = 0.57$. The tendency of a positive
correlation becomes visible.

Panels (d), (e) and (f) of Fig.~\ref{cor_elly} present the analogous
plots for the maps TH3 and H$\alpha$. This pair of data shows a high
degree of correlation: $r_p = 0.81$. Note that the plot in panel (e)
corresponding to $W(a, \vec x)$ at $a=1\farcm 0$ (a local maximum in
$r_w(a)$, see Fig.~\ref{cor_ha}) gives a better confined cloud of
points than the initial data in panel (d), in spite of the fact that
the correlation coefficient is $r_w = 0.71$ (less than $r_p$). This
means that the scatter in the points in panel (d) is compensated by the
larger number of points along the central strip of the cloud.
The points in
panel (e) are also more homogeneously distributed than in panel (d).
Panel (f) shows again the scale $a=2\farcm 8$, on which TH3 and
H$\alpha$ are less correlated than on the scale $a=1\farcm 0$
($\simeq1.6$~kpc, the typical width of a spiral arm).

When analysing galactic images, the scales which characterize the width
of the spiral arms are of special interest. Therefore we calculated the
mean value of $r_w$ for the range of scales $0\farcm 4 < a < 1\farcm
2$. These scales are also more certain in the sense that they are large
enough with respect to the size of the beam and small enough to give
satisfactory statistics over the area of the galactic image. The values
derived are given in the lower part of the Table~1 (below the diagonal).

Comparing the corresponding values of $r_p$ and $r_w$ in Table~1,
three groups of map pairs can be distinguished: 1) the two correlation
coefficients are similar (the difference between $r_p$ and $r_w$ is
less than 25\%), 2) the two correlation coefficients strongly differ,
$r_p/r_w \approx 2$ or even $r_w <0$ ($r_p$ is never negative
in the table), 3) comme ci - comme \c{c}a.

{\bf Group 1} includes most of the pairs containing $\HI$ or TH3 (only
the pairs $\HI$--PI6, TH3--PI6 strongly drop out of this rule) and
several pairs with H$\alpha$: H$\alpha$--R, H$\alpha$--I,
H$\alpha$--DUST. The coefficients $r_p$ and $r_w$ are practically
equal for $\HI$--I6, $\HI$--NTH3, TH3--H$\alpha$, and TH3--I6. In the
whole table only the pairs $\HI$--CO and $\HI$--DUST give $r_w >r_p$.

{\bf Group 2} includes first of all the correlations with PI6.
Three pairs (PI6--R, PI6--TH3, PI6--H$\alpha$) yield negative $r_w$.
Strong differences between $r_w$ and $r_p$ occur in R--NTH3
(0.29/0.86), R--CO (0.35/0.83), NTH3--CO (0.32/0.80), NTH3--H$\alpha$
(0.34/0.64), and NTH3-DUST (0.47/0.89).

\begin{figure*}
\centerline{\psfig{file=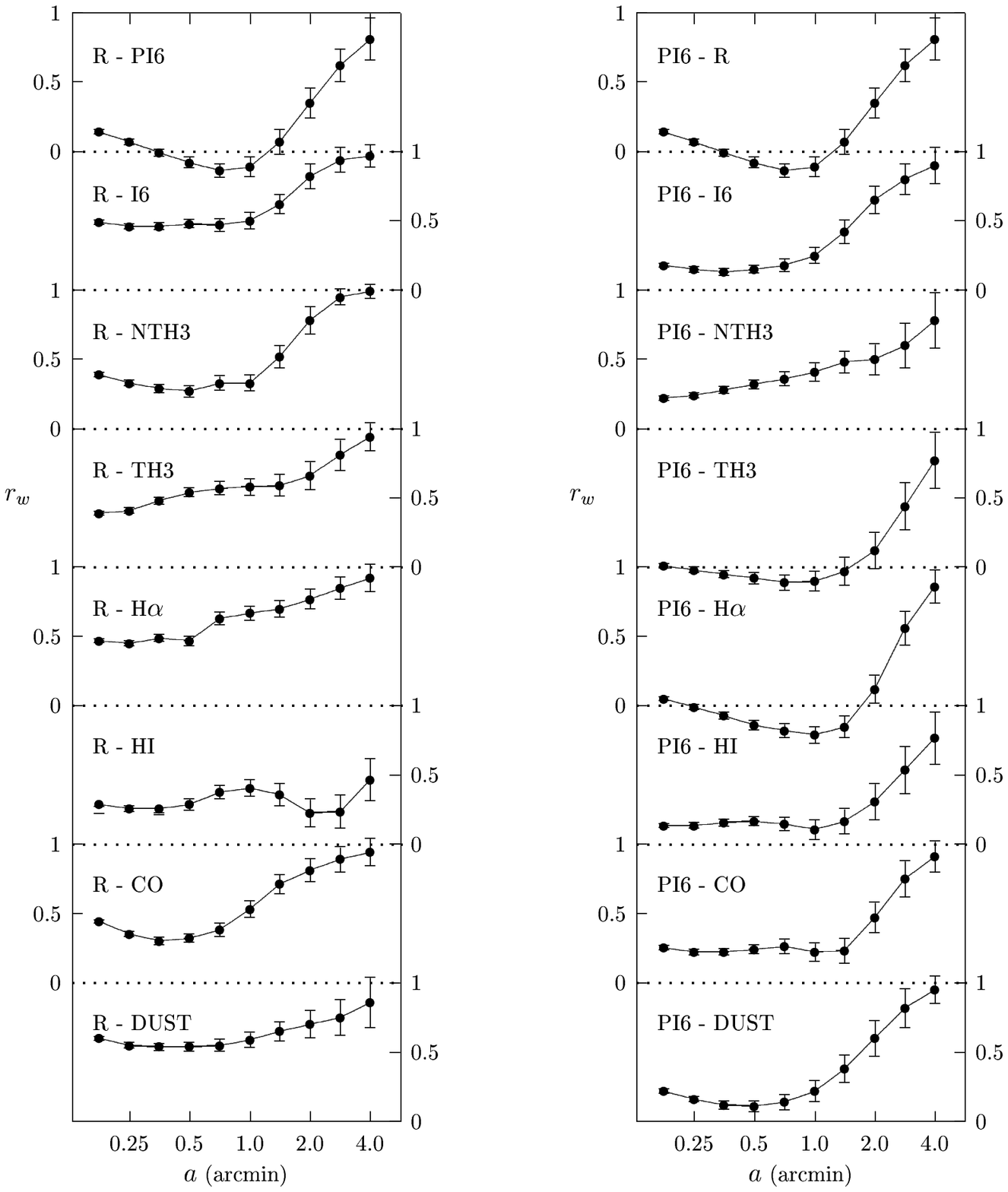,width=14truecm,%
       bbllx=90pt,bblly=152pt,bburx=544pt,bbury=689pt}}
\caption{NGC\,6946. Wavelet cross-correlations for red light (R, left) and
polarised radio emission (PI6, right) versus each of the other maps.}
\label{cor_red_pi}
\end{figure*}

We calculated 36 cross-correlation functions $r_w(a)$, one for each pair of
galactic maps. Because of space limitation we present the correlation
curves $r_w(a)$ only for two sets of pairs (see Fig.~\ref{cor_red_pi}:
red emission R (left column) and polarised emission PI6 (right column)
correlated with each of the other 8 maps.

For most pairs $r_w$ varies considerably with scale.
Only a few curves display a weak dependence on scale, for example
R--DUST, which leads to similar values for $r_w$ and $r_p$ (Group
1). The best coincidence of $r_w$ and $r_p$, however, is obtained
when $r_w$ has a {\it maximum} on the scales chosen for the calculation
of $r_w$ for Table~1. The pairs R--$\HI$ (see Fig.~\ref{cor_red_pi})
and H$\alpha$--TH3 (Fig.~\ref{cor_ha}) are good examples of this case.
Also the pairs $\HI$--CO, H$\alpha$--I6, TH3--DUST show well pronounced
local maxima on the scale of the arm width.

Obviously, the main part of Group 2 (high contrast between $r_p$
and $r_w$) consists of pairs for which the $r_w$ curve has a
{\it minimum} on intermediate scales. Apart from all pairs including
PI6, the pairs CO--NTH and CO--R belong to this group.
In these cases the classical cross-correlation coefficient
can be misleading (see Sect.~6.2).

\section{Conclusions and Discussion}

In this paper we have used the wavelet analysis to find the scales of
dominant structures in maps of the spiral galaxy NGC\,6946 observed
in various spectral ranges and to study cross-correlation coefficients
between pairs of maps. Our conclusions are summarised below.

\subsection{Wavelet spectra}

We would like to stress once again that the scaling analysis of
observational data requires an adequate choice of the mathematical
tool. We have shown that the structure function, which is a commonly
used method for scaling analysis of small-scale turbulence, is not
suitable for scales which are close to the dominating large-scale
structures (see Figs.~\ref{sp_6IM} and \ref{sp_HA}). Due to poor scale
resolution, the structure function may suggest that a continuous range
of scales with a power law exists (as e.g. in Beck et al. 1999)
even when this is not present in the data. On the contrary, the
traditional Fourier technique applied to real data gives
very spiky spectra in which the separation of real maxima and high
harmonics can be difficult.

We have shown that wavelets (provided the choice of the wavelet is
adequate to the aim of the analysis and to the analysed data) allow to
obtain reliable spectral characteristics of the object even if the
range of scales in the map is relatively small (about one and a half
decades in our cases). A suitable choice of wavelets leads to smooth
spectra which clearly indicate the dominant scales in the analysed data.

Using wavelets, the spectral characteristics of 9 optical, radio and
infrared maps of the galaxy NGC\,6946 were calculated. The wavelet
spectra in Fig.~\ref{spectra_myf} show three types of curves:

a) R, I6, CO and DUST increase smoothly with increasing scale. These
maps show structures on small and large scales (clouds, spiral
arms, an extended central region and extended diffuse emission).

b) The spectrum of PI6 is practically flat on all scales because
this map neither contains an extended central region nor extended
diffuse emission.

c) TH3 and H$\alpha$ show a prominent maximum on intermediate
scales corresponding to the width of the spiral arms.
The maximum in TH3 occurs on a somewhat smaller scale than
that in H$\alpha$. A possible explanation is the increase of
absorption of the H$\alpha$ emission on small scales where gas
densities are generally higher (see also Fig.~\ref{cor_TH}).

We conclude that the wavelet spectrum reveals signatures of different
phenomena in an object and allows to compare their relative importance
within the object as well as between different objects or wavelength
ranges.

Only two spectra, H$\alpha$ and thermal emission TH3, include an
increasing part on small scales, which (with great caution) can be
interpreted as an indication of the Kolmogorov inertial range in which
the energy is transferred from large to small scales. In
Fig.~\ref{spectra_myf} we show the expected theoretical slope of 5/3
as thin lines at $a<0\farcm 5$ (800~pc) for H$\alpha$ and at $a<0\farcm
35$ (560~pc) for TH3. We also analysed the spectrum of the H$\alpha$
image at full resolution ($\simeq 2\arcsec$) and found that the slope
of 5/3 continues down to a scale of about $0 \farcm 1$ (160~pc). Note
that this slope should characterize the two-dimensional projection of a
three-dimensional turbulent field with Kolmogorov spectral properties.

Some spectra in Fig.~\ref{spectra_myf} include an increasing part with
a slope of less than 1. Red light R shows a slope $\approx 0.6$ for
$0\farcm 25 \le a \le 1\arcmin$, in CO the slope is close to $0.7$ for
$0\farcm 25 \le a < 1\arcmin$, and in the same range
of scales the DUST emission is characterized by a slope $\approx 0.5$.
The total radio emission I6 shows a large range of scales with growing
energy, but the slope is not constant and depends on the choice of the
range; a weak maximum near $a=0\farcm 5$ may indicate two subintervals
with different slopes. The mean slope in the range $0\farcm 25 \le a
\le 1\farcm 4$ is about 0.6--0.7.

Structure functions with a slope of about 2/3 over a range of scales
have been interpreted as an indication of two-dimensional turbulence
(Minter \& Spangler 1996; Beck et al. 1999). This interpretation,
however, is uncertain because the physical behaviour of
two-dimensional turbulence is quite different from three-dimensional
turbulence.

In 2-D turbulence two inertial ranges are possible (Kraichnan
1967). This is due to an additional conservation law, i.e. in the
limit of small viscosity the total square of vorticity, called
{\it enstrophy\/}, becomes an integral of motion as well as the
total energy. The first inertial range, the range of inertial
transport of enstrophy to smaller scales, appears on scales
smaller than the scale where the turbulence is excited. The
dimensional arguments lead to a very steep spectrum $E(k)\sim
k^{-3}$, which corresponds to a structure function $S_2\sim l^2$.
No average {\it spectral} flux of energy is expected in this
inertial range. The second inertial range can appear on scales
larger than the scales on which the energy input into the flow
occurs. Although characterized by the usual Kolmogorov spectral
slope ($E(k)\sim k^{-5/3}$ or $S_2\sim l^{2/3}$), it has very
different physical properties because the energy in this inertial
range can be transferred only to {\it larger\/} scales (so-called
inverse energy cascade). It is important to note that the
formation of the inverse energy cascade is a relatively slow process
and implies a stationary forcing on intermediate scales (see, e.g.,
Borue (1994) or Babiano, Dubrulle \& Frick (1997)).
Two-dimensional turbulence with a slope of 2/3 implies the
existence of an inertial range of scales with inverse cascade.
This means that the energy of turbulent motion on these scales
does not emerge from the larger (galactic) scale, but is
transferred from smaller (sub-galactic) scales on which
nevertheless the motion already has to be two-dimensional, which
is improbable.

At the moment we cannot
explain the origin of the slopes of about 2/3 detected in the wavelet
spectrum of R, CO, DUST and I6 on small scales. Furthermore,
we do not understand why a slope of about 5/3 is observed only
for the ionised gas (H$\alpha$ and TH3). Future images with higher
resolution are needed in order to extend the analysis to smaller scales.

\subsection{Cross-correlations}

We have introduced a new cross-correlation characteristic based on
wavelet decomposition, named the {\it wavelet cross-correlation\/}, which
allows to analyse the correlation as a function of scale. The classical
cross-correlation coefficient is misleading if a bright, extended central
region or an extended disk exists in the galactic images.
In such cases (see R, I6, NTH3,
CO and DUST in Table~1) the classical cross-correlation is dominated by
the large-scale structure which can be identified as the galaxy's
disk with its general radial decrease in intensity (see
Fig.~\ref{red_slices},
bottom right), while the
(more interesting) correlation on scales of the spiral arms can be much
worse. In Table~1 this is clearly indicated by the wavelet correlation
coefficients $r_w$ which are much lower than the corresponding values
of the classical coefficient $r_p$.

In classical cross-correlation analysis, the general radial decrease
of a galaxy's intensity can be removed by dividing all values
within a certain radial range by its mean (see e.g. Hoernes
et al. 1998). This is similar to removing the largest scale from
a map only if the radial decrease is smooth and independent of
azimuthal angle. The wavelet technique is recommended as the
more reliable method.

\begin{figure}
\centerline{\psfig{file=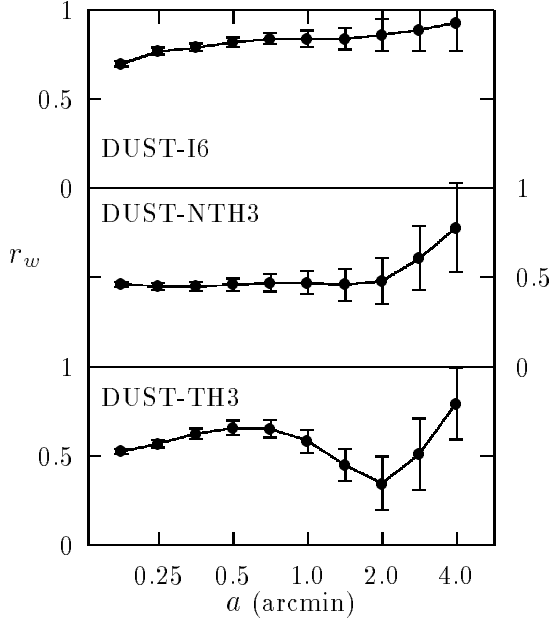,width=8truecm,%
       bbllx=90pt,bblly=250pt,bburx=310pt,bbury=500pt}}
\caption{NGC\,6946. Wavelet cross-correlations for mid-infrared dust
emission (DUST) versus total (I6), nonthermal (NTH3) and thermal (TH3)
radio emission.}
\label{cor_DUST}
\end{figure}

The best correlation, with $r_w >0.75$ for any scale $a$,
yields the pair DUST--I6 (Fig.~\ref{cor_DUST}).
The correlation between DUST and each of the two components of
radio continuum, nonthermal and thermal emission, is significantly
worse. This indicates that the correlation between the total
intensities is not due to a single process like star-formation
activity or coupling of magnetic fields to gas clouds, but needs a
cooperation of physical processes on all scales. A detailed
analysis will be the subject of a separate paper.

Similarly, the far-infrared emission of the total dust in M31 is highly
correlated with the total radio emission, while the correlations
between the components warm dust -- thermal radio and between
cool dust -- nonthermal radio are worse (Hoernes et al. 1998;
Berkhuijsen, Nieten \& Haas 2000).

We have shown that the {\it anticorrelation} between PI6 and TH3
(and between PI6 and H$\alpha$) can be quantified using
wavelet analysis. This phenomenon, which is due to the phase
shift between the magnetic arms and the optical arms
(Beck \& Hoernes 1996; Frick et al. 2000),
exists only on intermediate scales and is
clearly visible as a minimum in Fig.~\ref{cor_ha} (lower curve) and
Fig.~\ref{cor_red_pi} (right). The position of this minimum indicates
the scale on which the anticorrelation is strongest, i.e. $\simeq
1\farcm 0$ or 1.6~kpc at the distance of NGC\,6946, which is the typical
width of a spiral arm.

The absolute value of the wavelet correlation coefficient in these
minima is relatively small (0.1--0.25 only). This fact can be explained
twofold. Firstly, the structures which are responsible for this
anticorrelation (the arms) occupy only a limited part of the image.
Secondly, in contrast to the illustrative example given in
Fig.~\ref{r1}, real arms are more irregular in space and in scales.
This is why the position of the local extremum in the correlation curve
$r_w(a)$ can be informative even if the absolute value of $r_w(a)$ is
relatively small.

\begin{figure}
\centerline{\psfig{file=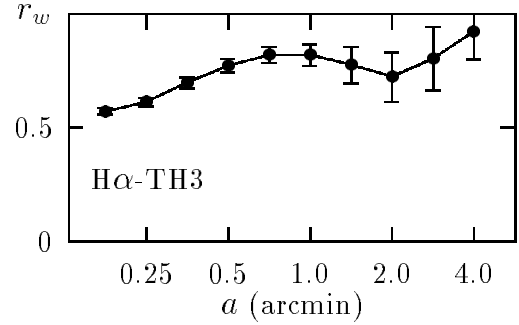,width=8truecm,%
       bbllx=100pt,bblly=310pt,bburx=300pt,bbury=440pt}}
\caption{NGC\,6946. Wavelet cross-correlation between
H$\alpha$ and thermal radio continuum emission (TH3).}
\label{cor_TH}
\end{figure}

Fig.~\ref{cor_TH} shows the correlation between the two
tracers of ionised gas. The correlation is high on intermediate
and large scales, but relatively low on small scales. This
gives further indication for enhanced absorption of the H$\alpha$
emission on small scales. Thus wavelet analysis can be used to
study the distribution of the absorbing dust.

Future radio images with better resolution will allow to
study the correlation on smaller scales.

\section*{Acknowledgments}
We thank Drs. Annette Ferguson and Daniel Dale for providing
H$\alpha$ and infrared maps of NGC\,6946 and Dr. Wolfgang Reich
for careful reading of the manuscript.
P.F. and I.P. acknowledge the MPIfR for hospitality. The visit
of P.F. to the MPIfR was supported by Deutscher Akademischer
Austausch Dienst (DAAD).

\label{lastpage}

\end{document}